\documentclass[11pt]{article}
\usepackage{graphicx,caption2,subfigure}
\makeatletter
\newcommand{\singlespacing}{\let\CS=\@currsize\renewcommand{\baselinestretch}{1.0}\tiny\CS}
\newcommand{\doublespacing}{\let\CS=\@currsize\renewcommand{\baselinestretch}{1.5}\tiny\CS}

\begin{document}
\title {On the Nature of the Rapidity-Spectra at RHIC and Some Other Energies.}
\author { Goutam
Sau$^1$\thanks{e-mail: sau$\_$goutam@yahoo.com}, S. K.
Biswas$^2$\thanks{e-mail: sunil$\_$biswas2004@yahoo.com},A. C. Das
Ghosh$^3$\thanks{e-mail: dasghosh@yahoo.co.in},
 A. Bhattacharya$^4$\thanks{e-mail: pampa@phys.jdvu.ac.in}$\&$ S.
Bhattacharyya$^5$\thanks{e-mail: bsubrata@www.isical.ac.in
(Communicating Author).}\\
{\small $^1$ Beramara RamChandrapur High School,}\\
 {\small South 24-Pgs,743609(WB),India.}\\
{\small $^2$ West Kodalia Adarsha Siksha Sadan,}\\
  {\small New Barrackpore, Kolkata-700131, India.}\\
  {\small $^3$ Department of Microbiology,}\\
    {\small Surendranath College,Kolkata-700009, India.}\\
   {\small $^4$ Department of Physics, }\\
 {\small  Jadavpur University, Kolkata- 700032, India.}\\
  {\small $^5$ Physics and Applied Mathematics Unit(PAMU),}\\
 {\small Indian Statistical Institute, Kolkata - 700108, India.}}
\date{}
\maketitle
\bigskip
\begin{abstract}
 On the basis of the Grand Combinational Model (GCM) outlined and
 somewhat detailed in the text, we have attempted to capture here
 the several interesting assorted characteristics of the
 rapidity-spectra of the major varieties of secondaries produced in
 diverse nuclear reactions at various energies, though the main
 thrust of our work lies on addressing the data-trends from RHIC-BNL
 experiments. Obviously the core of the present approach is purely
 phenomenological. Still, the method and the model address the
 features of the data modestly well. And the method appears to have
 the rich potentiality, if the systematic sets of data for
 rapidity-studies at gradually increasing energies were available.
\end{abstract}
\newpage
\doublespacing
\section{Introduction}
Unlike the cases of $p_T$-spectra, theoretical (even
phenomenological), studies on the rapidity-spectra are relatively
much fewer, though these two observables [($p_T$,y) or ($p_T,\eta$)]
are just twins for analyzing the nature of invariant cross-sections
for the secondary particles produced in the arranged laboratory
collisions at very high energies. In fact, the experimentalists also
are not immune from such biases, mainly due to the accompanying
difficulties in the measurement processes. And the net effect of all
these put together is the number of data sets on rapidity spectra is
obviously much less abundant and they suffer from large
measuremental uncertainties or large errors of both systematic and
statistical nature. Despite these strong and somewhat insurmountable
limitations, in the present work we will confine ourselves to the
studies on rapidity-spectra produced in Pb+Pb reactions at
marginally high energies and in RHIC-BNL interactions involving
Au+Au collisions at $\sqrt{s_{NN}}$ = 200 GeV in the light of Grand
Combinational Model (GCM). The stimulus stems from the fact that in
the past we analyzed a large part of the data on $p_T$-spectra for
RHIC-BNL collisions and a considerable part of the rapidity-spectra
at CERN SPS-energies. So, for the sake of completeness of our
studies on the behaviour of this specific model (GCM), we need to
check whether this model as such (or any extension of the model for
higher energies) could explain the relatively sparse data on the
rapidity-spectra and some other related properties involving these
experimental measurements on rapidity-dependent physical
observables.
\par The plan of this paper is as follows. In Section 2, we attempt
to provide an outline of the Grand Combinational Method (GCM) and
give the successive procedural steps. The Section 3 gives an inkling
into the basic root of the expressions used in Section2. The Section
4 exhibits the results and some discussions on them. The last
Section (Section 4) contains the final summary and the conclusive
remarks.
\section{The Working Model : An Outlook}
Following Faessler\cite{Faessler1}, Peitzmann\cite{Peitzmann1} and
also the work of Schmidt and Schukraft\cite{Schmidt1}, we propose
here a generalized empirical relationship between the inclusive
cross-section for production of any particle represented by $Q$ in
nucleus-nucleus collision and production of the same in
nucleon-nucleon collisions in the following form :
\begin{equation}
 E\frac{d^3\sigma}{dp^3}|_{AB \longrightarrow QX}  \sim (AB)^{\epsilon(y,p_T)}E\frac{d^3\sigma}{dp^3}|_{PP \longrightarrow QX} ,
\end{equation}
where $Q$ is the secondary, like pion/kaon/proton/antiproton,
produced in high energy nucleus-nucleus(AB) or in proton-proton(PP)
collisions. The term $\epsilon(y,p_T)$ could be expressed in the
factorization form $\epsilon(y,p_T)$ = $f(y)g(p_T)$. While
investigating a specific nature of dependence of the two variables
(y and $p_T$), either of them is assumed to remain averaged or with
definite values. Speaking in clearer terms, if and when rapidity
dependence is studied by experimental group, the transverse momentum
is integrated over certain limits and is absorbed in the
normalization factor. So the effective formula for rapidity spectra
turns into
\begin{equation}
 \frac{d\sigma}{dy}|_{AB \longrightarrow QX}  \sim (AB)^{f(y)}\frac{d\sigma}{dy}|_{PP \longrightarrow QX}.
\end{equation}
The main bulk of work, thus, converges to the making of an
appropriate choice of form for f(y). And the necessary choice is to
be made on the basis of certain premises and physical considerations
which do not violate the canons of high energy interactions.
\par Applying the concept of both limiting fragmentation and the Feynman Scaling hypothesis we attempt at providing
first a fit to the data on rapidity distributions for proton-proton
collisions at several $\sqrt{s}$ values, ranging from $\sqrt{s}$ =
23 GeV to 63 GeV with a parametrization by 3-parameter formula :
\begin{equation}
\frac{1}{\sigma}\frac{d\sigma}{dy}=C_1(1+\exp\frac{y-y_0}{\Delta})^{-1}
\end{equation}
where $C_1$ is a normalization constant and $y_0$, $\Delta$ are two
parameters. The choice of the above form made by Thome' et
al\cite{Thome1} was intended to describe conveniently the central
plateau and the fall-off in the fragmentation region by means of the
parameters $y_0$ and $\Delta$ respectively. For all five energies in
PP collisions the value of $\Delta$ was obtained to be $\sim$ 0.55
for pions\cite{De1} and kaons\cite{De2}, and $\sim$ 0.35 for
protons/antiprotons\cite{De2}. And these values of $\Delta$ are
generally assumed to remain the same in the ISR ranges of energy.
Still, for very high energies, and for direct fragmentation
processes which are quite feasible in very high energy heavy
nucleus-nucleus collisions, such parameter values do change somewhat
prominently, though in most cases with marginal high energies, we
have treated them as nearly constant.
\par Now, the fits for the rapidity (pseudorapidity)
 spectra for non-pion secondaries produced in the PP reactions at various energies are phenomenologically
 obtained by De and Bhattacharyya\cite{De2} through the making of
 suitable choices of $C_1$ and $y_0$. It is observed that for various secondaries the values of $y_0$
 remain almost constant and do not show up any sharp species-dependence amongst the secondaries.
  However, for pions it gradually increases with energies and the energy-dependence
 of $y_0$ is empirically proposed to be expressed by the relationship\cite{De1} :
\begin{equation}
y_0=0.55\ln\sqrt{s_{NN}}+0.88
\end{equation}
\par However, the energy-dependence of $y_0$ is studied here just for gaining insights in their nature
and for purposes of extrapolation to the various higher energies (in
the frame of $\sqrt{s_{NN}}$) for several nucleon-nucleus and
nucleus-nucleus collisions. The specific energy (in the c.m. system,
 $\sqrt{s_{NN}}$) for every nucleon-nucleus or nucleus-nucleus collision is first worked out by
 converting the laboratory energy value(s) in the required c.m. frame energy value(s). Thereafter
 the value of $y_0$ to be used for computations of inclusive cross-sections of nucleon-nucleon collisions
 at particular energies of interactions is extracted from Eq. (4) for corresponding obtained energies.
 This procedural step is followed for calculating the rapidity (pseudorapidity)-spectra for only the pions
 produced in nucleon-nucleus and nucleus-nucleus collisions\cite{De1}. But, for the studies on the rapidity-spectra
 of the non-pion secondaries produced in the same reactions one does neither have the opportunity to take recourse
 to such systematic step, nor could they actually resort to this rigorous procedure, as both the values of $y_0$ and
 $\Delta$ (as was done by De and Bhattacharyya\cite{De2}) do not depict any sensitive energy-dependences. So, we have considered
 only the average values of $y_0$ and $\Delta$ to study the rapidity spectra for non-pion secondaries produced in nucleon (nucleus)-nucleus collisions.
\par Our next step is to explore the nature of $f(y)$ which is envisaged to be given generally by a polynomial form noted
below :
\begin{equation}
f(y) = \alpha + \beta y + \gamma y^2,
\end{equation}
where $\alpha$, $\beta$ and $\gamma$ are the coefficients to be
chosen separately for each AB collisions (and also for AA collisions
when the projectile and the target are same). But, in so far as
dealing with the rapidity spectra is concerned, the above expression
suffers from a limitation which we would discuss later in some
detail in section 3.2. Besides, some other points are to be made
here. The suggested choice of form in expression (5) is not
altogether fortuitous. In fact, we got the clue from one of the
previous work by one of the authors (SB)\cite{Bhattacharyya1} here
pertaining to the studies on the behavior of the EMC effect related
to the lepto-nuclear collisions. In the recent past Hwa et
al\cite{Hwa1} also made use of this sort of relation in a somewhat
different context. Now we go back to our original discussion.
Combining Eqs. (2) and (5) the final working formula for
$\frac{dN}{dy}$ in various AB (or AA) collisions can be expressed by
the following relation :
\begin{equation}
\frac{dN}{dy}|_{AB \rightarrow QX} = C_2(AB)^{\alpha + \beta y + \gamma y^2}\frac{dN}{dy}|_{PP \rightarrow QX}
 = C_3(AB)^{\beta y + \gamma y^2}(1 + \exp \frac{y-y_0}{\Delta})^{-1},
\end{equation}
where $C_2$ is the normalization constant and $C_3$=$C_2(AB)^\alpha$
is another constant as $\alpha$ is also a constant for a specific
collision at a specific energy. The parameter values for different
nucleus-nucleus collisions are given in Table1 - Table 9. However,
in the next section (Section 3) we attempt at providing hints to the
possible origin of our very starting equation [Eqn. 1] in this
Section.
\section{Basic Sources of the Theoretical Framework : An Outline}
At the very root, our approach owes a lot to the basics of the
Glauber Model and to the pioneering developmental works by Gribov
who showed how to incorporate the Glauber Model for interactions of
hadrons with nuclei into the general framework of relativistic
quantum theory. Subsequently, an important contribution to the
theory multiparticle production was made by Gribov jointly with
Abramovsky and Kancheli which is normally referred to as the AGK
cutting rules. Later Capella et al\cite{Czyz1} combined in an
excellent work the outcomes of all of them in a concise paper from
which we pick up the expressions having very close resemblance with
both the physical ideas, the terms and also the mathematical
structure.
\par An important consequence of the space-time structure of the
scattering diagrams for hadron-nucleus interactions as studied by
Capella et al is : for inclusive cross sections all rescatterings
cancel with each other and the cross sections are determined solely
by the very common impulse approximation. This statement is valid
asymptotically in the central rapidity region (y $\simeq$ 0) and for
cases where the masses of the intermediate studies are both limited
and energy-independent.
\par The inclusive cross section for the production of a hadron i is expressed, for a given impact parameter
 b, in terms of inclusive cross section for hN
 interactions\cite{Capella1}
\begin{equation}
E \frac{d^3\sigma^i_{hA}(b)}{d^3p} = T_A(b) E \frac{d^3\sigma^i_{hN}}{d^3p}
\end{equation}
where $T_A(b)$ is the nuclear profile function ($\int d^2b T_A(b) =
A$). After integrating over b we get
\begin{equation}
E \frac{d^3\sigma^i_{hA}}{d^3p} = A E \frac{d^3\sigma^i_{hN}}{d^3p}
\end{equation}
\par The total and inelastic hA cross sections in the Glauber model can be easily calculated
and are given for heavy nuclei by well known expressions. For
example
\begin{equation}
\sigma^{in}_{hA} = \int d^2b(1-exp(-\sigma^{in}_{hN}T_A(b)))
\end{equation}
\par The situation for nucleus-nucleus collisions is much more complicated.
There are no analytic expressions in the Glauber model for
heavy-nuclei elastic scattering amplitudes.
 The problem stems from a complicated combinatorics and from the existence of dynamical correlations related
 to ``loop diagrams"\cite{Andreev1,Boreskov1}. Thus, usually optical-type approximation\cite{Formanek1,Czyz1}
 and probabilistic models for multiple rescatterings\cite{Pajares1} are used. For inclusive cross sections in
 AB-collisions the result of the Glauber approximation is very simple to formulate due to the AGK cancelation theorem.
 It is possible to prove, for an arbitrary number of interactions of nucleons of both nuclei\cite{Boreskov2},
 that all rescatterings cancel in the same way  as for hA-interactions. Thus a natural generalization of eq. (7) for
 inclusive spectra of hadrons produced in the central rapidity region in nucleus-nucleus interactions takes place in the
 Glauber approximation
\begin{equation}
E \frac{d^3\sigma^i_{AB}(b)}{d^3p} = T_{AB}(b)E \frac{d^3\sigma^i_{NN}}{d^3p}
\end{equation}
where $T_{AB}(b) = \int d^2s
T_A(\overrightarrow{s})T_B(\overrightarrow{b}-\overrightarrow{s})$.
After integration over b eq. (10) reads
\begin{equation}
E \frac{d^3\sigma^i_{AB}}{d^3p} = AB E \frac{d^3\sigma^i_{NN}}{d^3p}
\end{equation}
\par It is to be noted here that eqs. (10), (11) are valid for an arbitrary set of Glauber diagrams and are not
influenced by the problem of summation of ``loop" diagrams mentioned above.
 \par The densities of charged particles can be obtained from eqs. (10), (11) by dividing
 them by the total inelastic cross section of nucleus-nucleus interaction. For example
 \begin{equation}
 \frac{dn^{ch}_{AB}(b)}{dy} = \frac{T_{AB}(b)}{\sigma^{in}_{AB}}\frac{d\sigma^{ch}_{NN}}{dy}
 \end{equation}
 and
 \begin{equation}
 \frac{dn^{ch}_{AB}}{dy} = \frac{AB}{\sigma^{in}_{AB}}\frac{d\sigma^{ch}_{NN}}{dy}
 \end{equation}
 In the following we shall use these results to calculate particle densities in the central rapidity region
 at energies of RHIC and LHC. It is to be seen that our starting
 expression [eqn. (1)] is just a generalization of the expression
 (11) shown here.
\section{Results}
This section is to be divided into three Subsections. In the first
one we summarize the procedural steps that follow from the
Section-2. And in the rest the results are delivered.
\subsection{}
The procedural steps for arriving at the results could be summed up
as follows :
\par (i) We assume that the inclusive
cross section (I.C.) of any particle in a nucleus-nucleus (AB)
collision can be obtained from the production of the same in
nucleon-nucleon collisions by multiplying it (I.C.) a product of the
atomic numbers of each of the colliding nuclei raised to a
particular function, $\epsilon(y,p_T)$, which at first is
unspecified (Equation 1).
\par (ii) Secondly, we accept that factorization of the function
$\epsilon(y,p_T)$ = $f(y)g(p_T)$ which helps us to perform the
integral over $p_T$ (Equation 2) in a relatively simpler manner.
\par (iii) Thirdly, we assume
a particular 3-parameter form for the pp cross section with the
parameters $C_1$, $y_0$ and $\Delta$ (Equation 3).
\par (iv) Finally,
we accept the ansatz that the function f(y) can be modeled by a
quadratic function with the parameters $\alpha$, $\beta$ and
$\gamma$ (Equation 5).
\subsection{}
 The results are shown here by the graphical plots with the accompanying tables for the parameter
values. In Fig.1 we draw the rapidity-density of the negative pions
for symmetric Pb+Pb collisions at several energies which have been
appropriately labeled at the top-right corner. In this context some
comments are in order. Though the figure represents the case for
production of negative pions, we do not anticipate and/or expect any
strong charge-dependence of the results. Besides, the \textbf{solid}
curves in all cases - almost without any exception - demonstrate our
GCM-based results. Secondly, the data on rapidity-spectra for some
high-energy collisions are, at times, available for both positive
and negative y-values. This gives rise to a problem in our method.
It is evident here in this work that we are concerned with only
symmetric collisions wherein the colliding nuclei must be identical.
But in our expression (6) the coefficient $\beta$ multiplies a term
which is proportional to y and so is \textbf{not} symmetric under
y$\rightarrow$(-y). In order to overcome this difficulty we would
introduce here $\beta$=0 for a large number of the graphical plots.
These plots are represented by Fig.1, Fig.2, Fig.5, Fig.6 and Fig.7
for the various secondaries produced in Au+Au/Pb+Pb collisions under
different conditions of relatively low c.m. energy values. The
parameter values in this particular case are presented in the Tables
1,2,3,6,7,8,9. The Fig.2a and 2b (both for $\beta$=0) are for
production of positive kaon and negative kaons in Pb+Pb interactions
at 30A GeV and 20A GeV. But the Fig.3 is based on the data on
pion-production in Au+Au collisions at several relatively low
energies for which data are available on dn/dy (rapidity density of
multiplicity) versus only positive y-values. The parameter values
are presented in Table4. And the diagrams in Fig.4 represent the
production of the major varieties of the secondaries produced in
Au+Au collisions at RHIC; herein too the data are obtained for only
positive values of rapidity. The corresponding parameter values are
presented in Table5.
\subsection{}
But the problem starts as soon as we attempt to reproduce the trends
in data simultaneously for both +ve and -ve rapidity values. The
graphs grow asymmetric even for symmetric collision. In Fig.5 with
$K^\pm$-meson varieties we tried to fix and focus the problem
encountered. The four labeled diagrams for $K^\pm$-meson varieties
shown in the present Fig.5 provide fits to the data with unchanged
$\Delta$-values ($\Delta$=0.55) for both positive and negative
y-values quite separately. The parameter values are to be obtained
from Table6 and Table7. These plots provide very important clues to
our understanding the nature of $\Delta$-values at very, very high
energies. For both kaon-antikaons one observes quite clearly that
the plots end up on the X-(rapidity)-axis for positive high-y-values
within the range +5 to +6 (positive) y-values, whereas for the
\textbf{negative} high y-values the spread widens to y-values
ranging between -6 to -7. This shows up somewhat convincingly to us
that the asymmetry problem arising in Fig.5 towards the +ve y-values
cannot be addressed properly and remedied without inserting
necessary changes in the accompanying sets of parameters, as the
treated energy-ranges for these Au+Au collision cases are very high
compared to the various sets of other data presented here. Our
studied diagnosis of the root of the problem forced us to alter the
previous $\Delta$-values with phenomenological induction of
increasing the nature of $\Delta$ with rising energy as was the case
for $y_0$. And, thus, making use of this ansatz, we were able to
redeem the symmetry-feature of the rapidity plots. The subsequent
two figures, the Fig.6 and Fig.7 strikingly illustrate how the
qualitative improvement in the nature of the plots on
rapidity-distributions for Au+Au reaction at RHIC was achieved by
changing $\Delta$-values. The parameter values for Fig.6 and Fig.7
are given in Table8 and Table9. The Fig.6 and Fig.7 (for $\beta$=0)
describe the rapidity spectra on the various non-pion secondaries as
labeled in Figs. 6(a), 6(b), 6(c), 6(d) and Figs. 7(a), 7(b), 7(c),
7(d) for both positive and negative y-values in Au+Au collisions at
RHIC energies with $\Delta$-value=1.70 and $\Delta$-value=3.5
respectively. Due to the paucity of data, we are not going to
present, at this juncture, any quantitative prediction on the nature
of rise of $\Delta$, though we qualitatively accept and use it in
this work.

\subsection{}
The diagrams shown in Fig.8(a) and Fig.8(b) represent the
model-based results on $K^+/\pi^+$ and $K^-/\pi^-$ respectively.
These plots are drawn on the basis of the figures shown in Fig.4
with the fit-parameters given in Table5. The data-trends have been
captured by them in a modestly right manner. The next plot (Fig.9)
pertains to the particle/antiparticle ratio-values which are really
quite crucial. This plot is drawn with the help of the figures shown
in Fig.4 and the Table5, that is, they were worked out on the basis
of the graphs drawn in Fig.4 and the accompanying table displayed in
the Table5.
\par
The rest two plots, Fig.10(a) and Fig.10(b), are drawn on the basis
of these revised figures with the gradually changed values of
$\Delta$. As no data on a charge-dependent basis for pions produced
in Au+Au reactions at RHIC energies with negative y-values were
available, we could not depict the $\pi^-/\pi^+$ ratio-behaviour in
Fig.10(a) and Fig.10(b) with changed $\Delta$-values. And the
physical justifications for introduction of such enhancements in the
values of $\Delta$ had already been hinted at quite emphatically in
Section 2. They also reproduce the features of the data, in a fairly
satisfactory manner. In fact, with changed $\Delta$-values they
describe, in our opinion, data-characteristics in a somewhat better
way.
\section{Discussion and Conclusions}
Surveying the model-based plots and judging by the degree and nature
of agreement between the measured data and the obtained results, we
could fairly sum up by a reasonable statement that the model(s)
applied here describe the data in a modestly satisfactory manner.
The used models have both a deductive origin and some
phenomenological components, with some conceptual physical strands :
(i) the results for the nucleus-nucleus collisions could be arrived
at by inducting the results of nucleon-nucleon collisions and some
relevant phenomenological product terms incorporating the physics of
nuclear dependences; (ii) the property of factorization is also
embedded in the results in an ingrained manner; (iii) there is a
certain degree of universality of nature among the secondaries
produced in nucleon-nucleon, nucleon-nucleus and nucleus-nucleus
interactions at high energies, by virtue of which the entire ranges
of rapidity-spectra have here been analyzed with the help of two
parameters ($y_0$ and $\Delta$) for PP collisions, and two other
($\beta$ and $y$) parameters for nucleus-nucleus collisions. Of
them, $y_0$ for pion-production cases alone could be obtained by a
formula [$eqn.$(4)] suggested by us. Similar behaviours of $y_0$ and
$\Delta$ are perfectly possible at very very high energies for all
non-pion secondaries as well, though, in most cases, we leave them
practically unchanged. However, There are some exceptions for Fig.6
and Fig.7. Thus, we are left with only two arbitrary parameters,
which one should try to relate with the physics of nuclear geometry
and collision dynamics. But there are some problems with the studies
of the rapidity-spectra : firstly, the systematic studies on the
rapidity spectra - both experimental and
theoretical/phenomenological - are comparatively quite fewer in
number ; secondly, the measurements suffer relatively from a higher
degree of uncertainty. In fact, the centrality-dependent studies on
rapidity spectra are not yet available, for which the studies on the
relationship between nuclear geometry and the rapidity-spectra
cannot yet be taken up either by us or others. Viewed from the angle
of these existing limitations and constraints, we could humbly claim
that the task of explaining the data-trends and behaviours on
rapidity spectra is done here with a modest degree of success. We
have checked that with the changed values of $\Delta$, it is easier
to describe the totality of data assembled here on several symmetric
collisions at various energies within the domain of high energy
ranges.
\par In our work, we have categorically pointed out
the problem of treating the cases of negative rapidity-values with
the help of the generalized expression given by expression (6),
wherein a specific term as is indicated clearly in 4.2 constitutes
the limitation of the method. For some other similar cases, however,
the method remains fully valid and works phenomenologically quite
well.
\par There is yet another drawback of this model. Since the
coefficients are fit for different species and collision energies we
cannot right now extrapolate to different collision energies or
connect the various coefficients, nor could we make a reliable
prediction for Pb+Pb collision at LHC on the basis of our fits to
the RHIC data for Au+Au collisions. As a result, the model lacks, so
far, any strong predictive power. And this is true of almost any
phenomenological model whatsoever. But this difficulty could be
remedied to a considerable extent, had there been availability of
reliable and accurate data at different high energies with
reasonably spaced intervals for both PP interactions and/or any
other symmetric high energy nuclear collisions. Besides, we are yet
to find and work out the ways and means for adaptation of this
methodology to any non-symmetric collision.

\begin{center}
\par{\textbf{Acknowledgements}}
\end{center}
\par
The authors express their thankful gratitude to the learned Referee
for some inspiring comments and constructive queries-cum-suggestions
for improving the quality of an earlier version of the manuscript.
\newpage
 \singlespacing

\newpage
{\singlespacing{
\begin{table}
\begin{center}
\begin{small}
\caption{Values of different parameters for production of
negative-pions in Pb+Pb collisions at different energies(for $\beta$
= 0) for both +ve and -ve rapidities.[Reference Fig. No.1]}
\begin{tabular}{|c|c|c|c|}\hline
 $Energy(GeV)$ & $C_3$ & $\gamma$ & $\frac{\chi^2}{ndf}$\\
 \hline
 $6.3$ & $87.426 \pm0.403 $  & $-0.042 \pm0.0006$ & $11.449/13 $ \\
 \hline
  $7.6$ & $100.095 \pm0.291 $ & $-0.036 \pm0.0003 $& $10.810/11 $ \\
 \hline
 $8.7$ & $ 118.278\pm0.274 $ & $-0.036 \pm0.0002 $& $2.943/5 $ \\
 \hline
  $12.3$ & $147.107 \pm0.401 $ & $-0.026 \pm0.0002 $& $10.218/6 $ \\
 \hline
  $17.3$ & $184.850 \pm0.449 $  & $-0.023 \pm0.0001 $& $3.637/4 $\\
 \hline
\end{tabular}
\end{small}
\end{center}

\begin{center}
\begin{small}
\caption{Values of different parameters for production of kaons in
Pb+Pb collisions at different energies for both +ve and -ve
rapidities (for $\beta$=0).[Reference Fig. No.2(a)]}
\begin{tabular}{|c|c|c|c|}\hline
 $Energy$ & $C_3$  & $\gamma$ & $\frac{\chi^2}{ndf}$\\
 \hline
 $20A GeV$ & $5.798\pm0.015 $  & $-0.0903 \pm0.0006$ & $0.673/3 $ \\
 \hline
  $30A GeV$ & $8.000 \pm0.027 $  & $-0.0726 \pm0.0005 $& $4.225/5 $ \\
 \hline
\end{tabular}
\end{small}
\end{center}

\begin{center}
\begin{small}
\caption{Values of different parameters for production of anti-kaons
in Pb+Pb collisions at different energies for both +ve and -ve
rapidities (for $\beta$=0).[Reference Fig. No.2(b)]}
\begin{tabular}{|c|c|c|c|}\hline
 $Energy$ & $C_3$ & $\gamma$ & $\frac{\chi^2}{ndf}$\\
 \hline
 $20A GeV$ & $16.990\pm0.059 $ & $-0.0430 \pm0.0020$ & $6.888/8 $ \\
 \hline
  $30A GeV$ & $21.589 \pm0.066 $  & $-0.0404 \pm0.0015 $& $7.282/11 $ \\
 \hline
\end{tabular}
\end{small}
\end{center}
\end{table}

\begin{table}
\begin{center}
\begin{small}
\caption{Values of different parameters for production of
negative-pions in Au+Au collisions at different energies for +ve
rapidities alone (for $\beta$$\neq$0).[Reference Fig. No.3]}
\begin{tabular}{|c|c|c|c|c|}\hline
 $Energy(GeV)$ & $C_3$ & $\beta$ & $\gamma$ & $\frac{\chi^2}{ndf}$\\
 \hline
 $8.8$ & $110.243\pm0.184 $ &$0.013\pm0.0002$ & $-0.032 \pm0.0002$ & $2.676/11 $ \\
 \hline
  $12.4$ & $147.023 \pm0.369 $ & $0.005\pm0.0002$ & $-0.019 \pm0.0001 $& $6.494/9 $ \\
 \hline
  $17.3$ & $185.589 \pm0.252 $ & $-0.006\pm0.0001$ & $-0.009 \pm0.0008 $& $3.238/8 $ \\
 \hline
\end{tabular}
\end{small}
\end{center}

\begin{center}
\begin{small}
\caption{Values of the different parameters for production of pions,
kaons, proton-antiprotons in Au+Au collisions at $\sqrt{S_{NN}}$ =
200GeV for +ve rapidities alone (for $\beta$$\neq$0).[Reference Fig.
No.4]}
\begin{tabular}{|c|c|c|c|c|}\hline
 $Production$ & $C_3$ & $\beta$ & $\gamma$ & $\frac{\chi^2}{ndf}$\\
 \hline
 $\pi^+$ & $303.480\pm0.400 $ &$0.0009\pm0.00007$ & $0.0052 \pm0.00002$ & $0.301/4 $ \\
 \hline
  $\pi^-$ & $300.938 \pm1.981 $ & $0.0009\pm0.0003$ & $0.0057\pm0.0001 $& $4.668/4 $ \\
 \hline
 $K^+$ & $49.169\pm0.099 $ &$0.0011\pm0.00009$ & $0.0049 \pm0.00003$ & $0.073/4 $ \\
 \hline
  $K^-$ & $46.996 \pm0.073 $ & $0.0010\pm0.00009$ & $0.0027\pm0.00003 $& $0.026/3 $ \\
 \hline
  $P$ & $24.862\pm0.659 $ &$0.009\pm0.002$ & $0.0059 \pm0.001$ & $3.431/8 $ \\
 \hline
  $\overline{P}$ & $17.794 \pm0.290 $ & $0.011\pm0.003$ & $-0.0059\pm0.001 $& $0.667/6 $ \\
 \hline
\end{tabular}
\end{small}
\end{center}
\end{table}

\begin{table}
\begin{center}
\begin{small}
\caption{Values of different parameters for production of kaon and
antikaon in Au+Au collisions at $\sqrt{S_{NN}}$ = 200GeV (for
$\beta$ = 0) for +ve rapidity only, taking $\Delta$=0.55.[Reference
Fig. No.5]}
\begin{tabular}{|c|c|c|c|}\hline
 $Production$ & $C_3$ & $\gamma$ & $\frac{\chi^2}{ndf}$\\
 \hline
 $Kaon$ & $ 47.128\pm0.051 $  & $-0.005 \pm0.0001 $& $1.941/09 $ \\
 \hline
  $Anti-kaon$ & $45.207 \pm0.064 $ & $-0.008 \pm0.0001 $& $1.018/06 $ \\
 \hline
 \end{tabular}
\end{small}
\end{center}
\end{table}

\begin{table}
\begin{center}
\begin{small}
\caption{Values of different parameters for production of kaon and
antikaon in Au+Au collisions at $\sqrt{S_{NN}}$ = 200GeV (for
$\beta$ = 0) for -ve rapidity only, taking $\Delta$=0.55.[Reference
Fig. No.5]}
\begin{tabular}{|c|c|c|c|}\hline
 $Production$ & $C_3$ & $\gamma$ & $\frac{\chi^2}{ndf}$\\
 \hline
 $Kaon$ & $ 47.808\pm0.076 $  & $-0.008 \pm0.0001 $& $1.990/09 $ \\
 \hline
  $Anti-kaon$ & $46.705 \pm0.069 $ & $-0.011 \pm0.0001 $& $3.461/08 $ \\
 \hline
 \end{tabular}
\end{small}
\end{center}
\end{table}

\begin{table}
\begin{center}
\begin{small}
\caption{Values of different parameters for production of proton,
antiproton, kaon and antikaon in Au+Au collisions at $\sqrt{S_{NN}}$
= 200GeV (for $\beta$ = 0) for both +ve and -ve rapidities, taking
$\Delta$=1.7.[Reference Fig. No.6]}
\begin{tabular}{|c|c|c|c|}\hline
 $Production$ & $C_3$ & $\gamma$ & $\frac{\chi^2}{ndf}$\\
 \hline
 $Proton$ & $49.770\pm0.690$  & $-0.011 \pm0.002$ & $2.083/09 $ \\
 \hline
  $Anti-proton$ & $21.103 \pm0.408 $ & $-0.016 \pm0.001 $& $4.636/13 $ \\
 \hline
 $Kaon$ & $ 55.524\pm0.111 $  & $-0.009 \pm0.0003 $& $2.867/04 $ \\
 \hline
  $Anti-kaon$ & $48.660 \pm0.109 $ & $-0.010 \pm0.0003 $& $3.859/05 $ \\
 \hline
 \end{tabular}
\end{small}
\end{center}
\end{table}

\begin{table}
\begin{center}
\begin{small}
\caption{Values of different parameters for production of proton,
antiproton, kaon and antikaon in Au+Au collisions at $\sqrt{S_{NN}}$
= 200GeV (for $\beta$ = 0) for both +ve and -ve rapidities, taking
$\Delta$=3.5.[Reference Fig. No.7]}
\begin{tabular}{|c|c|c|c|}\hline
 $Production$ & $C_3$ & $\gamma$ & $\frac{\chi^2}{ndf}$\\
 \hline
 $Proton$ & $59.529\pm1.028$  & $-0.009 \pm0.001$ & $4.924/13 $ \\
 \hline
  $Anti-proton$ & $25.107 \pm0.469 $ & $-0.016 \pm0.001 $& $5.160/13 $ \\
 \hline
 $Kaon$ & $67.962\pm0.228 $  & $-0.011 \pm0.0003 $& $18.708/09 $ \\
 \hline
  $Anti-kaon$ & $66.694 \pm0.198 $ & $-0.013 \pm0.0003 $& $9.524/05 $ \\
 \hline
 \end{tabular}
\end{small}
\end{center}
\end{table}

\newpage
\begin{figure}
\centering
\includegraphics[width=2.5in]{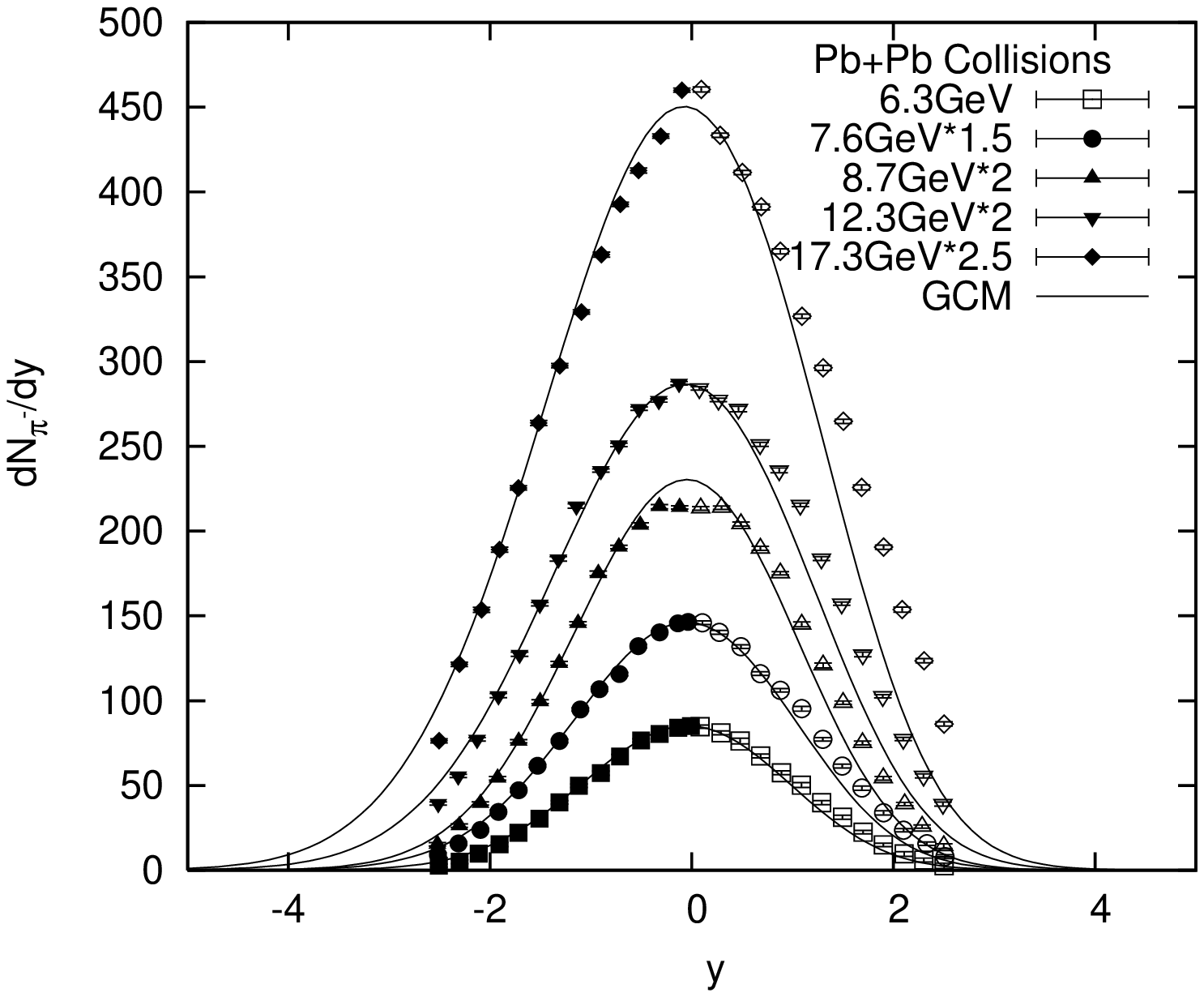}
\caption{Rapidity spectra for secondary pions produced in Pb+Pb interactions at different energies for $\beta$ =0. The different
experimental points are from Ref.\cite{Mitrovski1} and the parameter values are taken from Table1.
The solid curves provide the GCM-based results.}

\subfigure[]{
\begin{minipage}{.5\textwidth}
\centering
\includegraphics[width=2.5in]{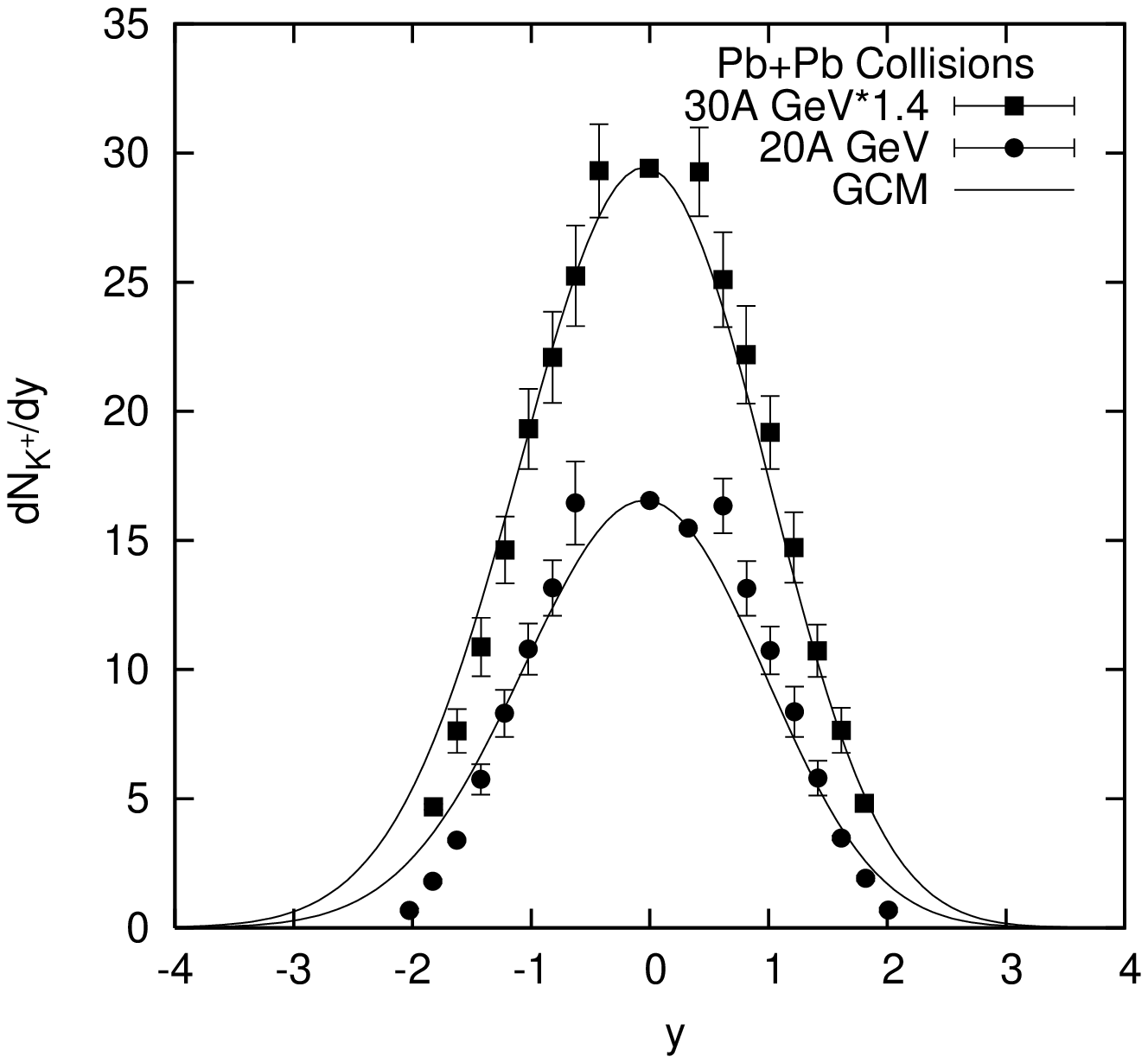}
\end{minipage}}%
\subfigure[]{
\begin{minipage}{.5\textwidth}
\centering
 \includegraphics[width=2.5in]{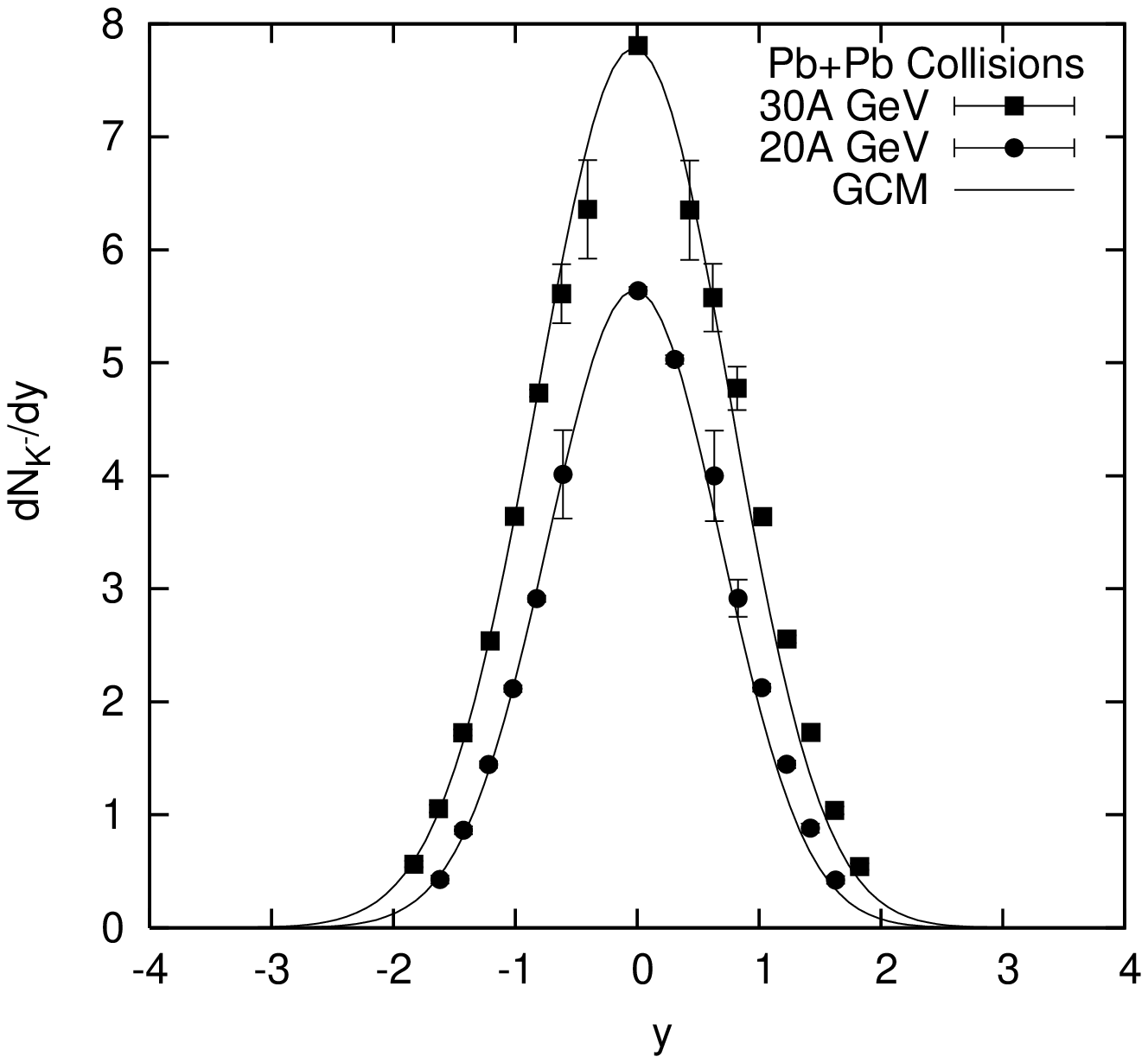}
 \end{minipage}}%
\caption{Rapidity distributions of $K^+$ and $K^-$ produced in
central Pb+Pb collisions at 20A and 30A GeV(both for $\beta$ = 0).
The different experimental points are from Ref.\cite{Alt1} and the
parameter values are taken from
 Table2 and Table3. The solid curves provide the GCM-based results.}
\end{figure}

\begin{figure}
\centering
\includegraphics[width=2.5in]{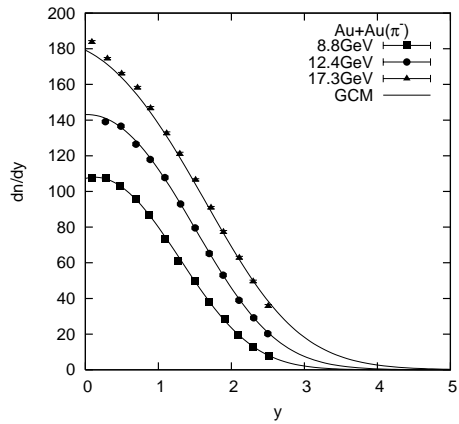}
\caption{Rapidity spectra for negative secondary pions produced in Au+Au interactions at different energies
 (for $\beta$$\neq$0). The different experimental points are from Ref.\cite{Wong1} and the parameter values are
  taken from Table4. The solid curves show the GCM-based results.}
\end{figure}

\begin{figure}
\subfigure[]{
\begin{minipage}{.5\textwidth}
\centering
 \includegraphics[width=2.5in]{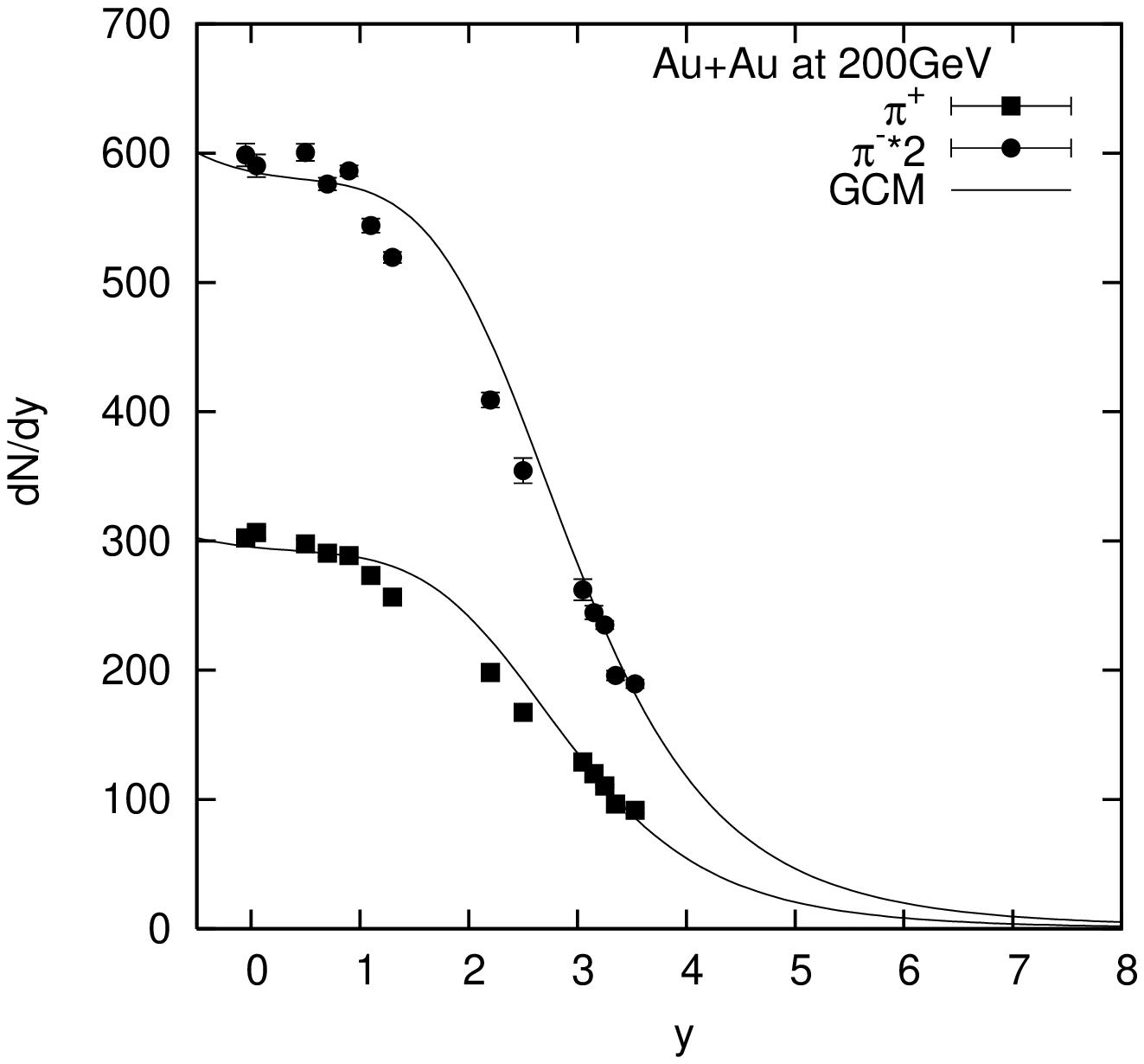}
\end{minipage}}%
\subfigure[]{
\begin{minipage}{0.5\textwidth}
  \centering
\includegraphics[width=2.5in]{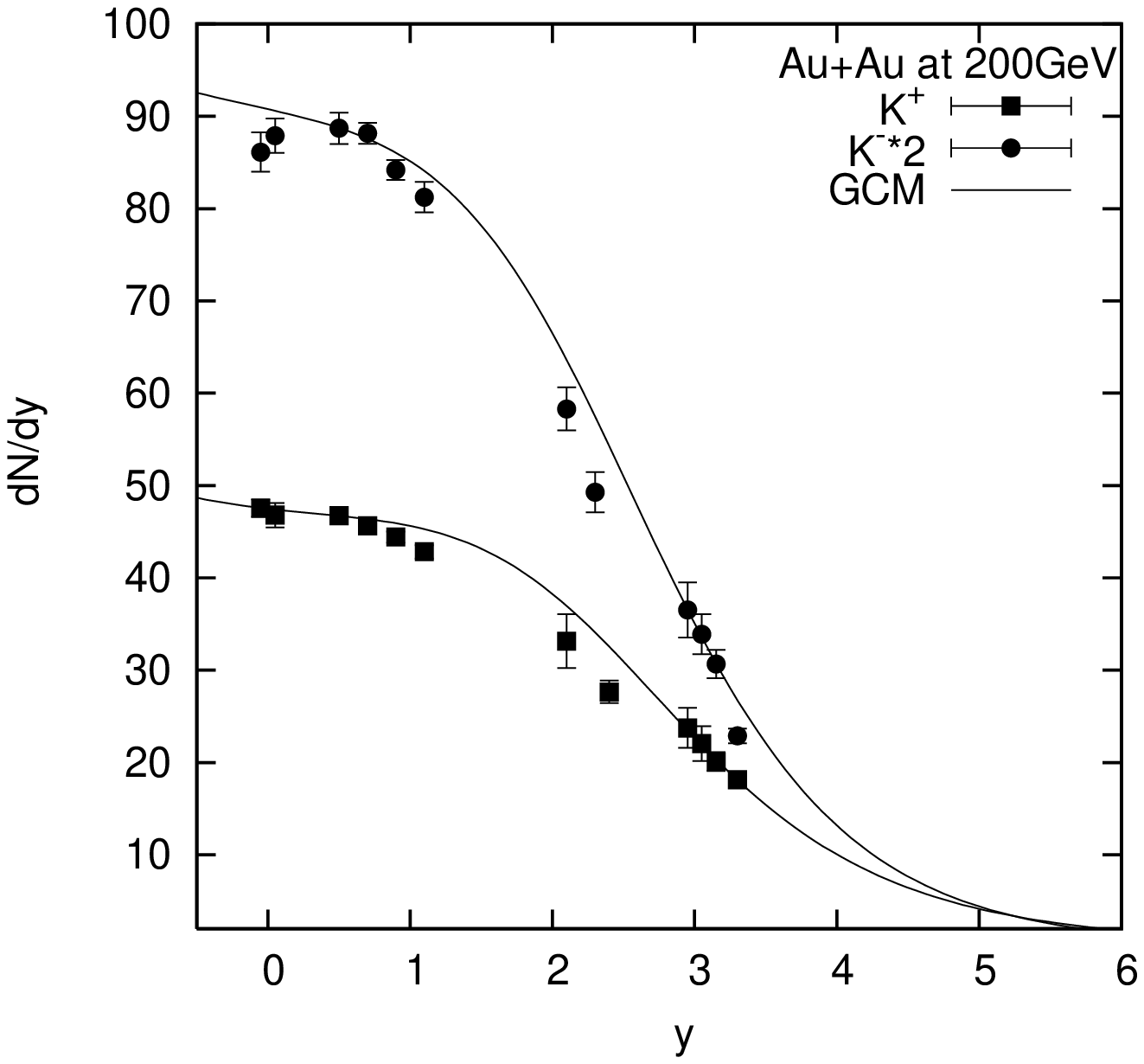}
\end{minipage}}%
\vspace{.01in} \subfigure[]{
\begin{minipage}{1\textwidth}
\centering
 \includegraphics[width=2.5in]{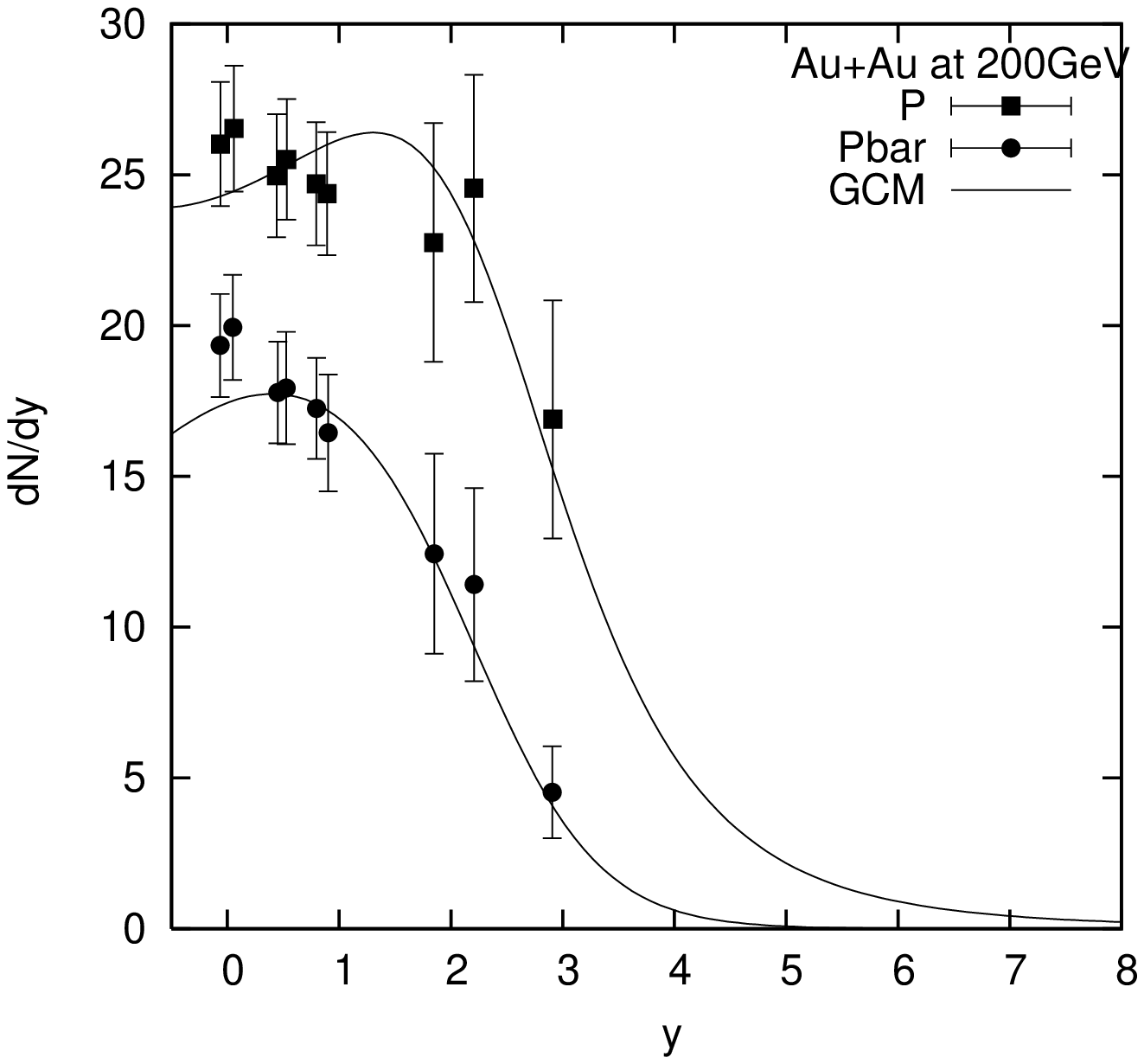}
\end{minipage}}%
\caption{Plot of $\frac{dN}{dy}$ vs. y for secondary pion, kaon, proton and antiproton produced in Au+Au interactions at
$\sqrt{S_{NN}}$ = 200GeV (for $\beta$$\neq$0). The different
experimental points are from Ref.\cite{Song1} and the parameter values are taken from Table5.
The solid curves provide the GCM-based results.}
\end{figure}

\begin{figure}
\subfigure[]{
\begin{minipage}{.5\textwidth}
\centering
\includegraphics[width=2.5in]{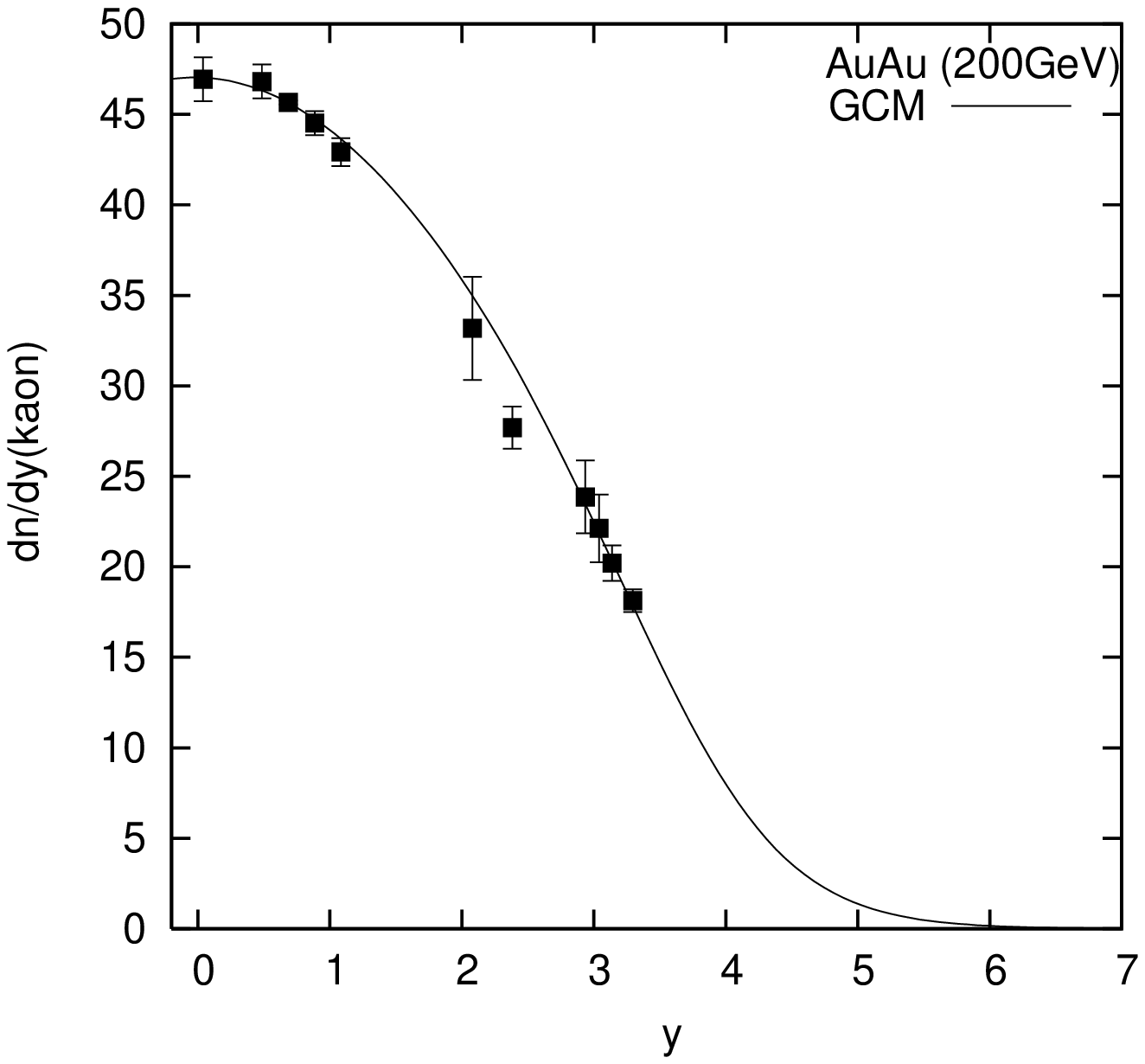}
\setcaptionwidth{2.6in}
\end{minipage}}%
\subfigure[]{
\begin{minipage}{0.5\textwidth}
\centering
 \includegraphics[width=2.5in]{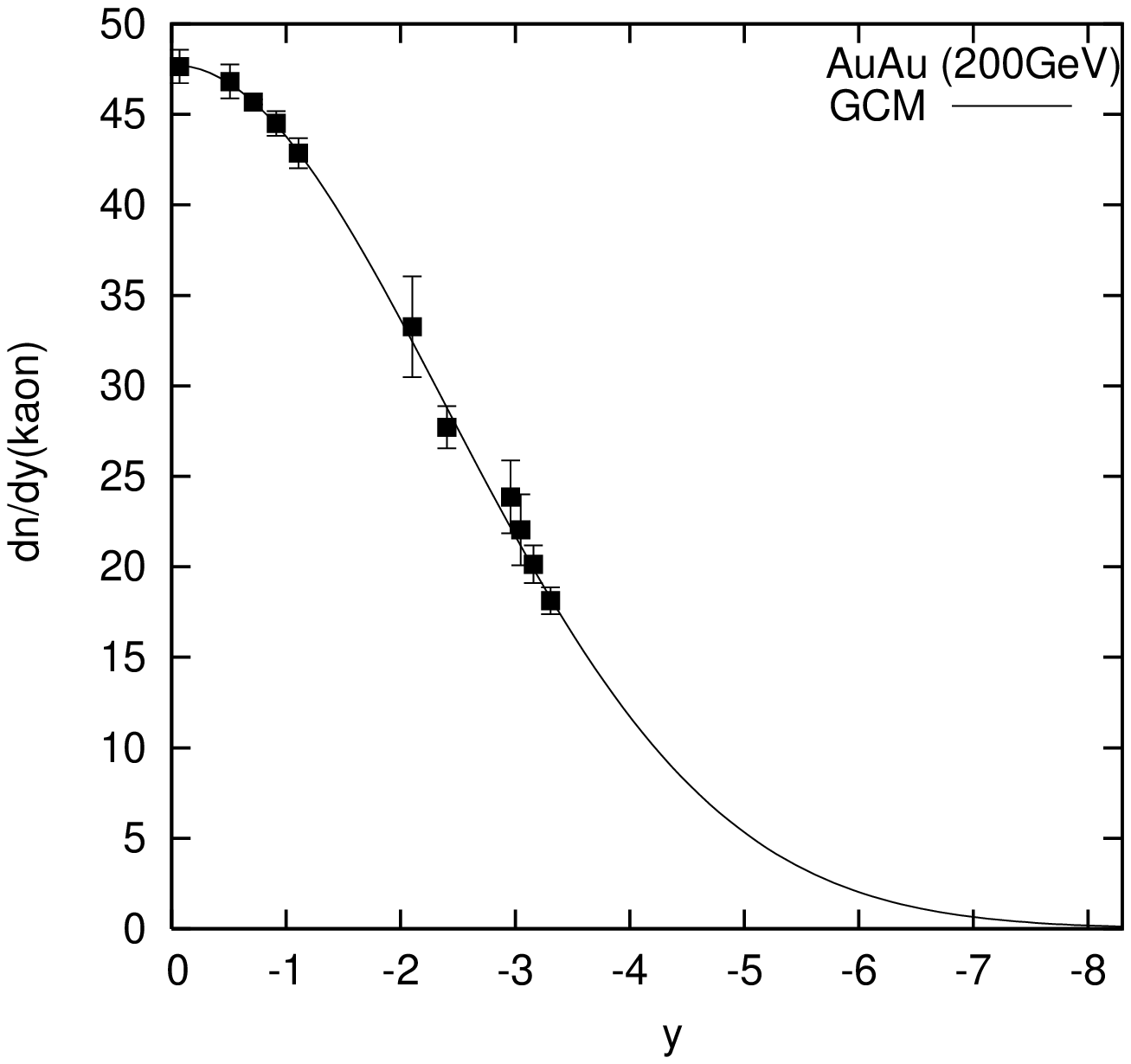}
 \end{minipage}}%
\vspace{0.01in} \subfigure[]{
\begin{minipage}{0.5\textwidth}
\centering
\includegraphics[width=2.5in]{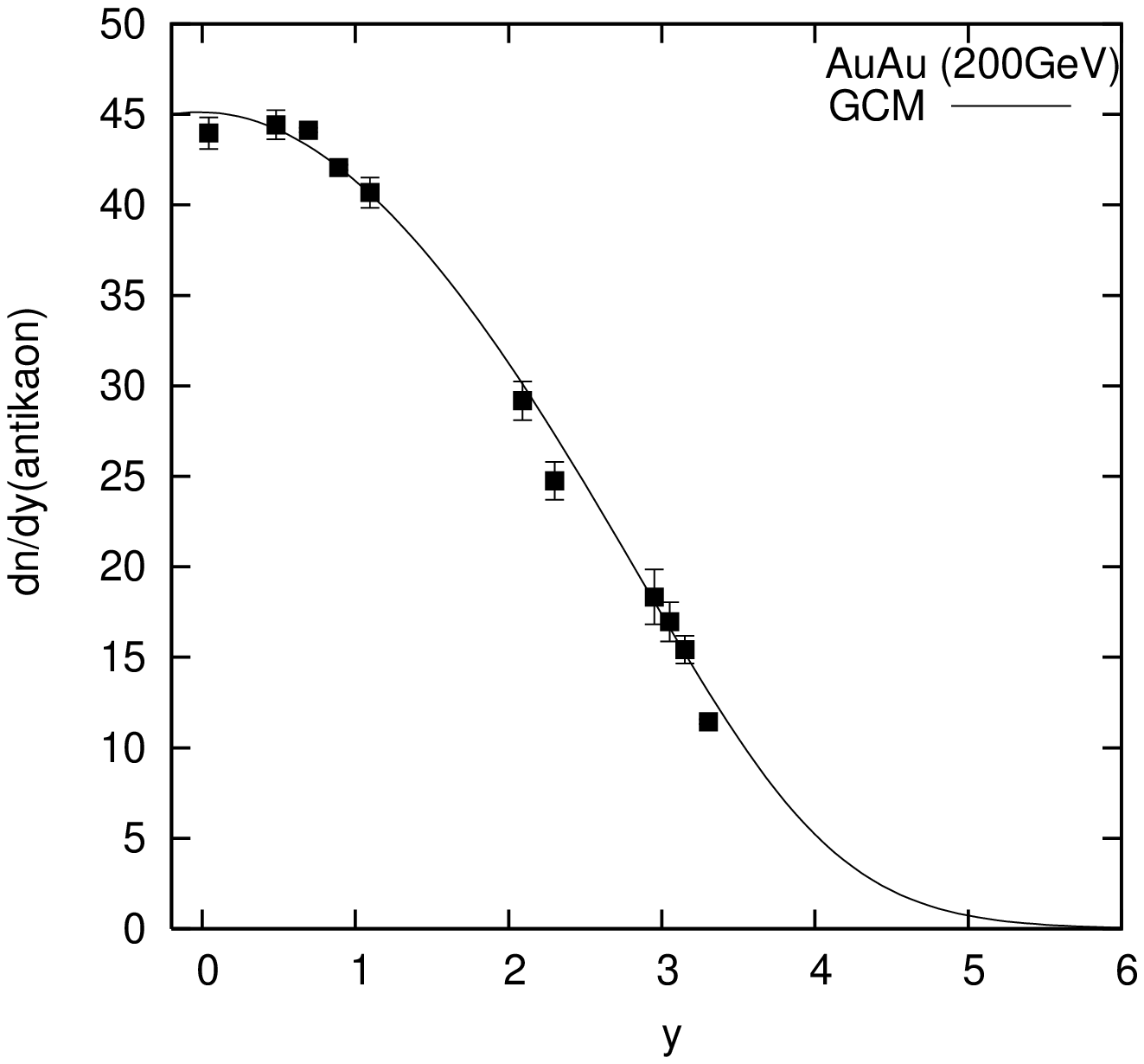}
\end{minipage}}%
\subfigure[]{
\begin{minipage}{.5\textwidth}
\centering
 \includegraphics[width=2.5in]{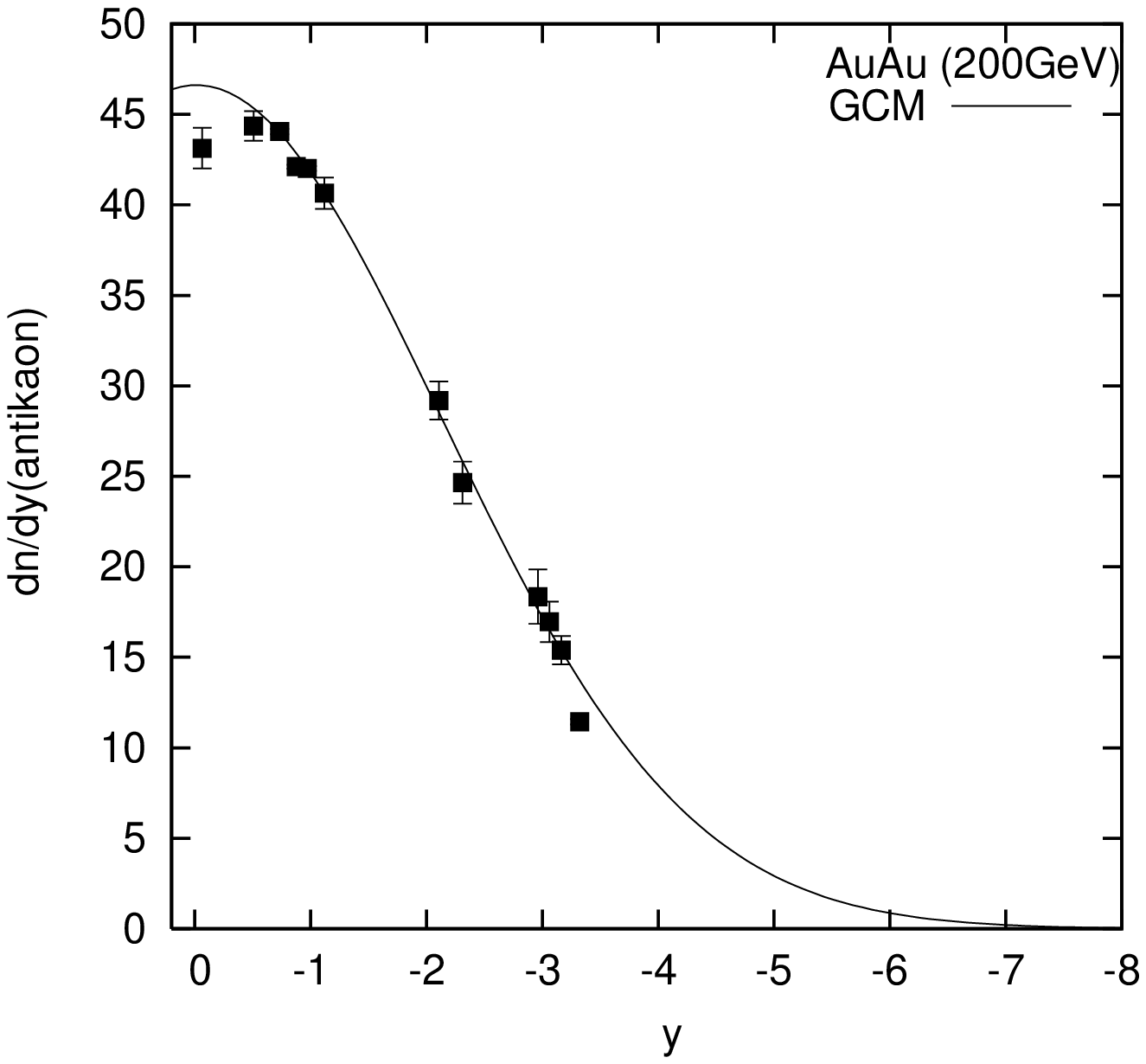}
 \end{minipage}}%
\caption{Plot of $\frac{dN}{dy}$ vs. y for secondary kaon and antikaon produced in Au+Au interactions for
positive and negative y-values separately at $\sqrt{S_{NN}}$ = 200GeV (for $\beta$=0), taking $\Delta$=0.55. The different
experimental points are from Ref.\cite{Uddin2} and the parameter values are taken from Table6 and Table7.
The solid curves provide the GCM-based results.}
\end{figure}

\begin{figure}
\subfigure[]{
\begin{minipage}{.5\textwidth}
\centering
\includegraphics[width=2.5in]{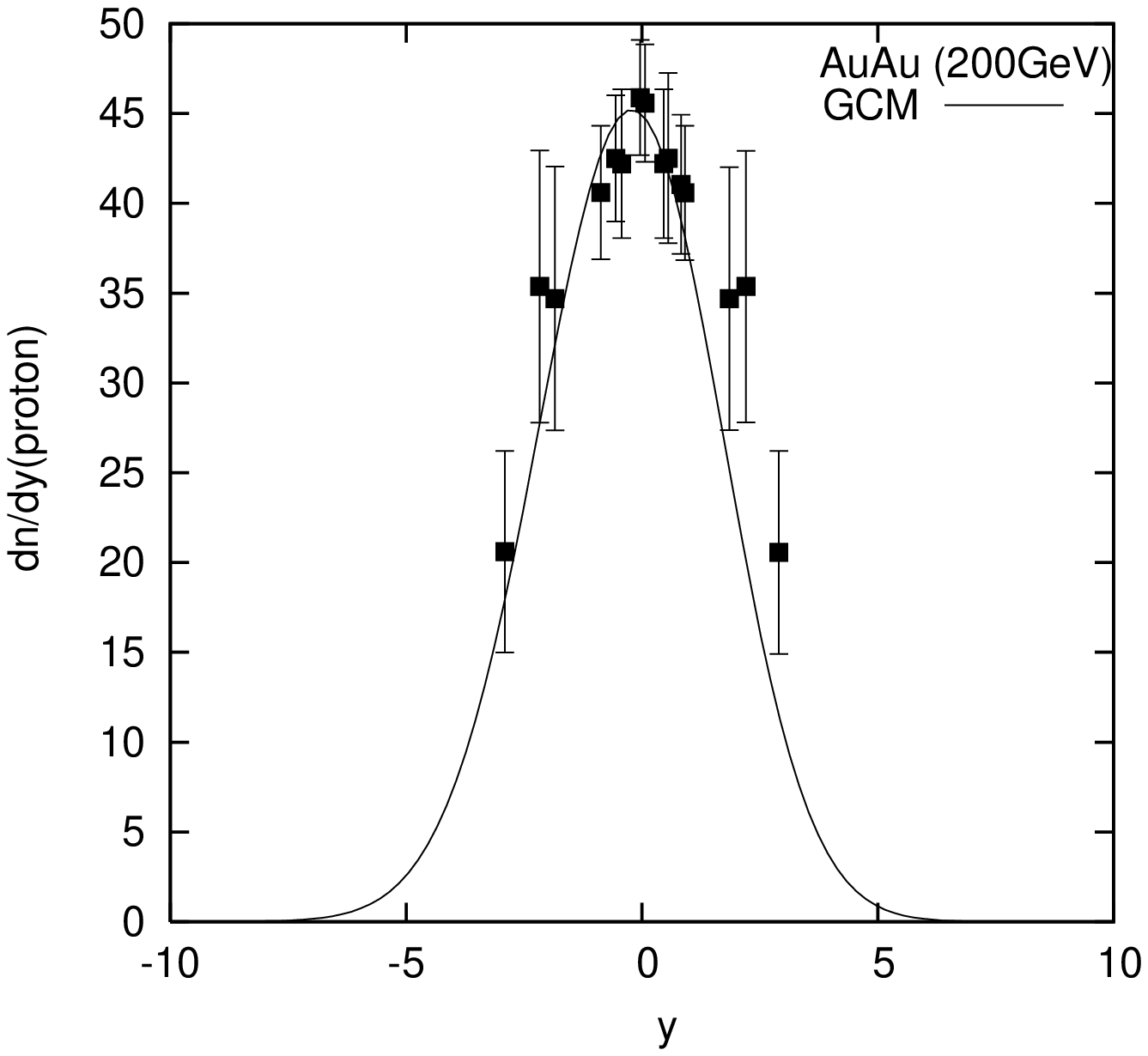}
\setcaptionwidth{2.6in}
\end{minipage}}%
\subfigure[]{
\begin{minipage}{0.5\textwidth}
\centering
 \includegraphics[width=2.5in]{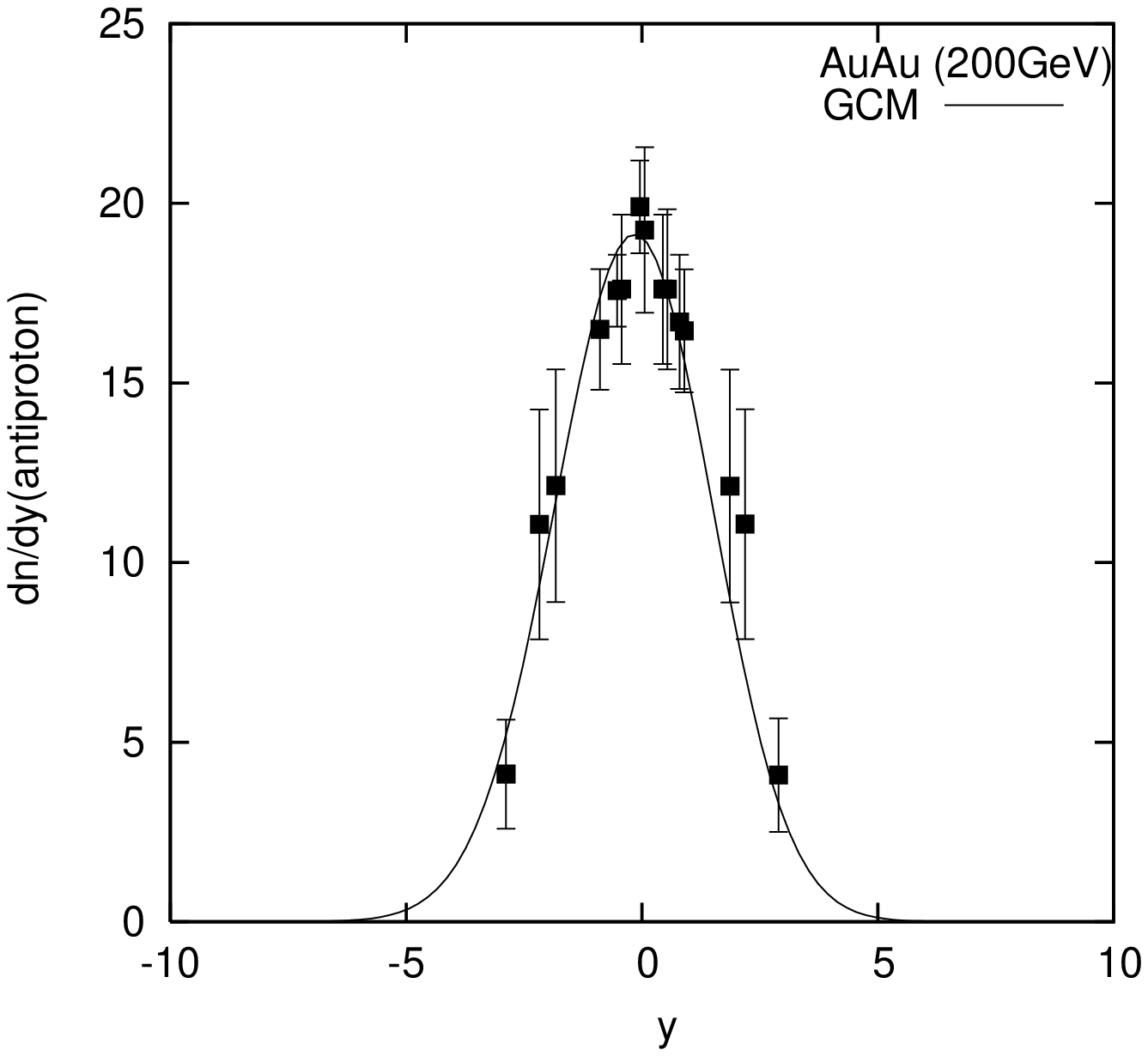}
 \end{minipage}}%
\vspace{0.01in} \subfigure[]{
\begin{minipage}{0.5\textwidth}
\centering
\includegraphics[width=2.5in]{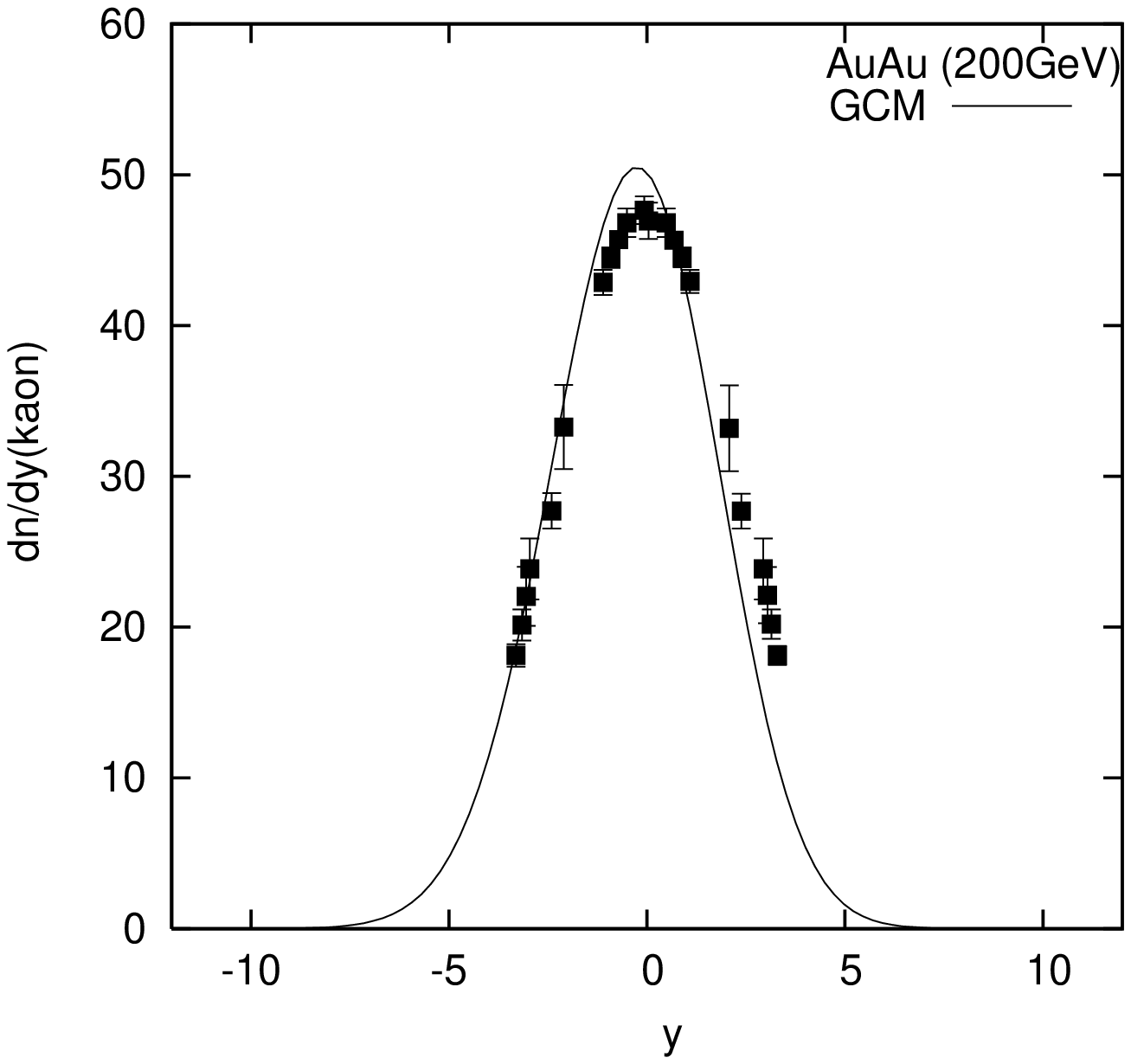}
\end{minipage}}%
\subfigure[]{
\begin{minipage}{.5\textwidth}
\centering
 \includegraphics[width=2.5in]{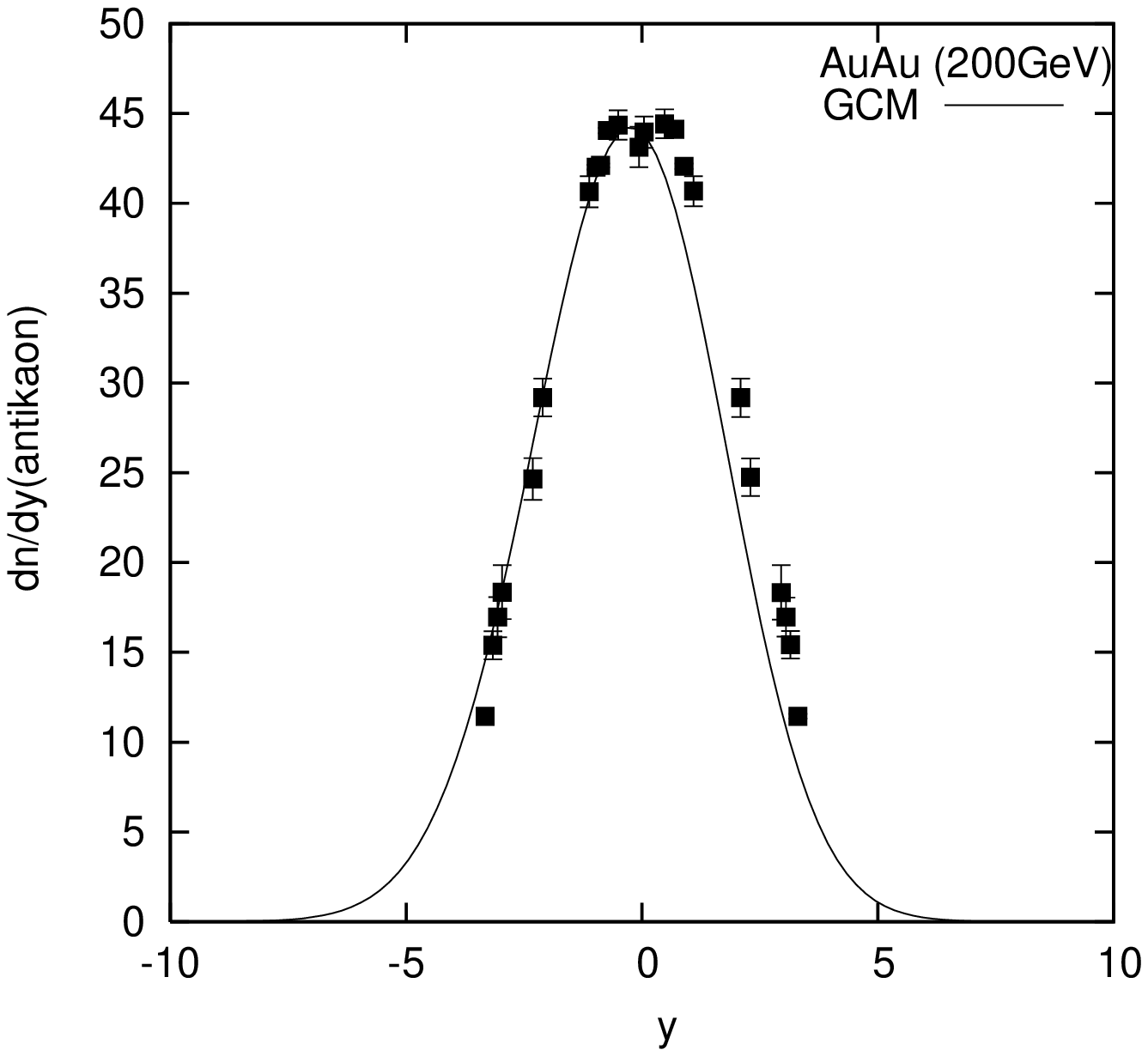}
 \end{minipage}}%
\caption{Plot of $\frac{dN}{dy}$ vs. y for secondary proton, antiproton, kaon and antikaon produced in Au+Au interactions for
positive and negative y-values separately at $\sqrt{S_{NN}}$ = 200GeV (for $\beta$=0), taking $\Delta$=1.7. The different
experimental points are from Ref.\cite{Uddin2,Uddin1} and the parameter values are taken from Table8.
The solid curves provide the GCM-based results.}
\end{figure}

\begin{figure}
\subfigure[]{
\begin{minipage}{.5\textwidth}
\centering
\includegraphics[width=2.5in]{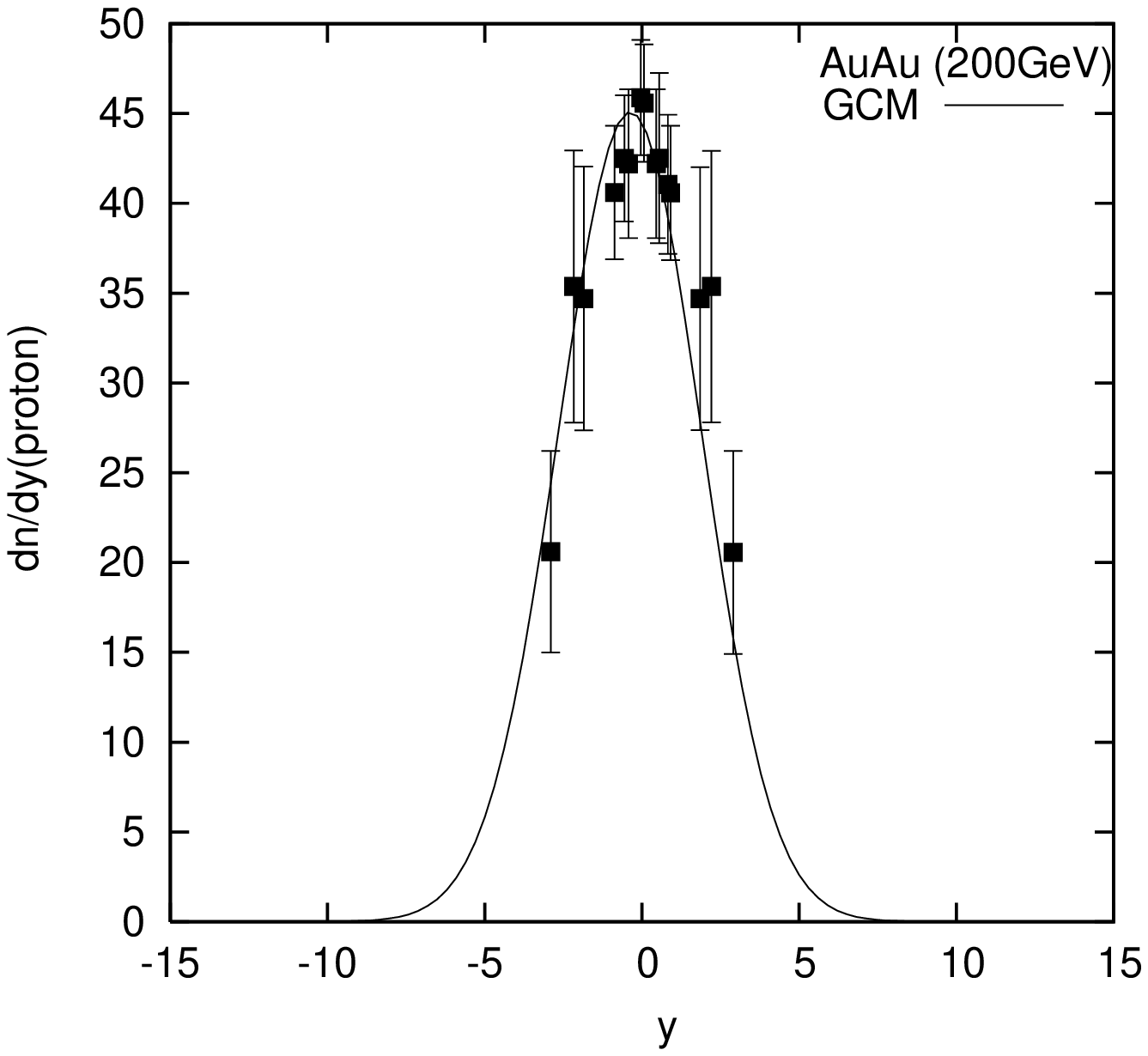}
\setcaptionwidth{2.6in}
\end{minipage}}%
\subfigure[]{
\begin{minipage}{0.5\textwidth}
\centering
 \includegraphics[width=2.5in]{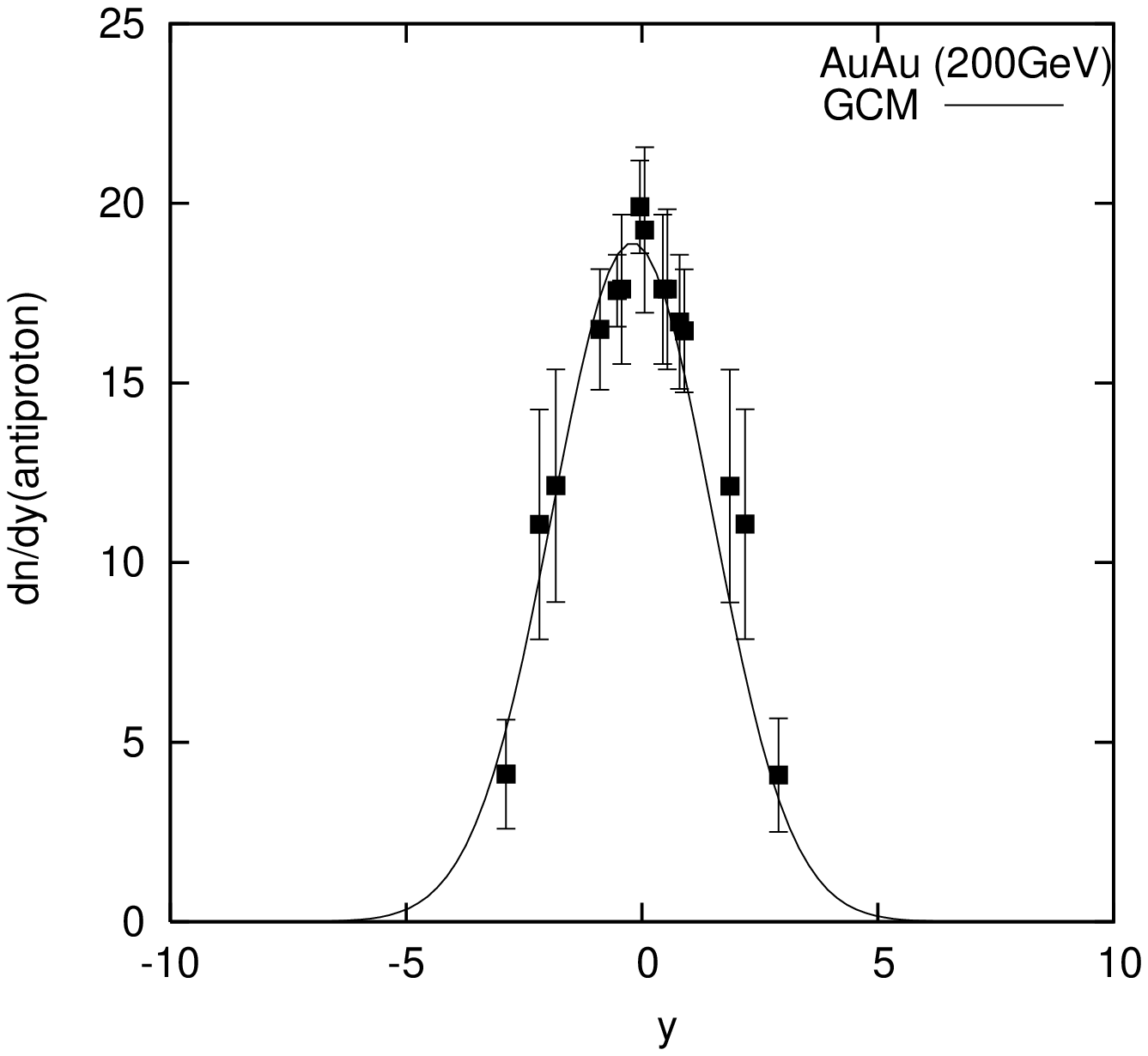}
 \end{minipage}}%
\vspace{0.01in} \subfigure[]{
\begin{minipage}{0.5\textwidth}
\centering
\includegraphics[width=2.5in]{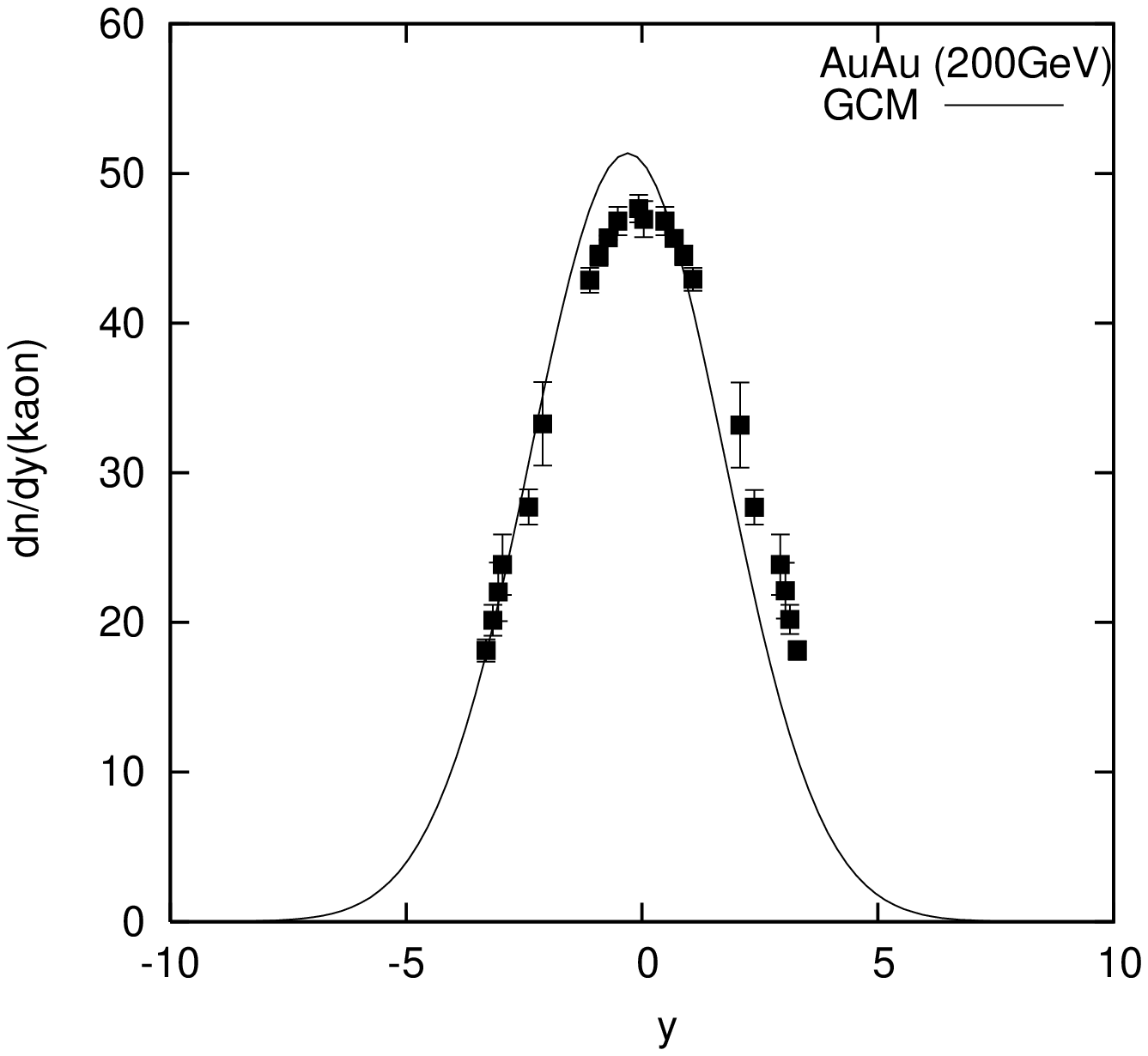}
\end{minipage}}%
\subfigure[]{
\begin{minipage}{.5\textwidth}
\centering
 \includegraphics[width=2.5in]{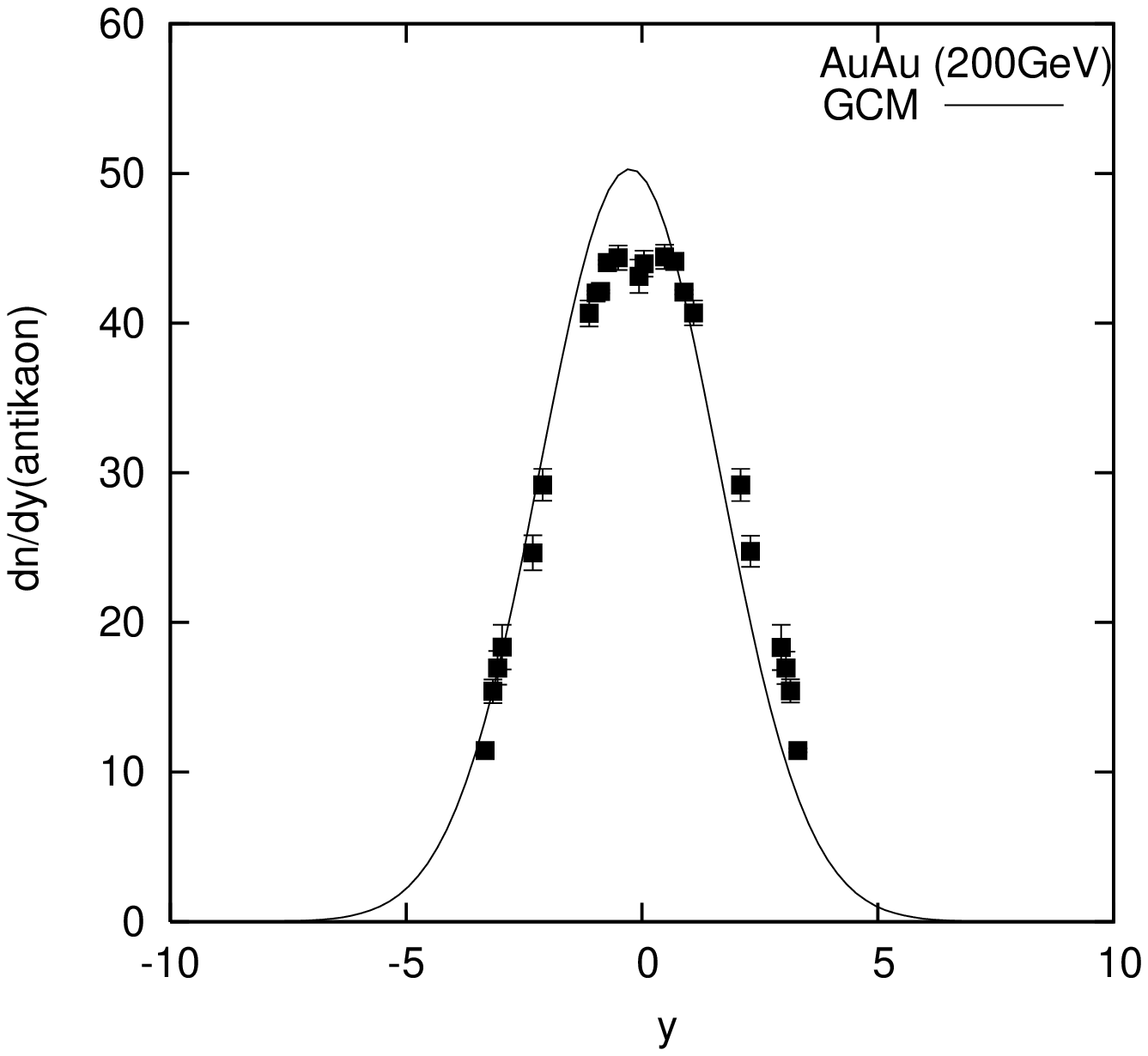}
 \end{minipage}}%
\caption{Plot of $\frac{dN}{dy}$ vs. y for secondary proton, antiproton, kaon and antikaon produced in Au+Au interactions for
positive and negative y-values separately at $\sqrt{S_{NN}}$ = 200GeV (for $\beta$=0), taking $\Delta$=3.5. The different
experimental points are from Ref.\cite{Uddin2,Uddin1} and the parameter values are taken from Table9.
The solid curves provide the GCM-based results.}
\end{figure}

\begin{figure}
\subfigure[]{
\begin{minipage}{.5\textwidth}
\centering
\includegraphics[width=2.5in]{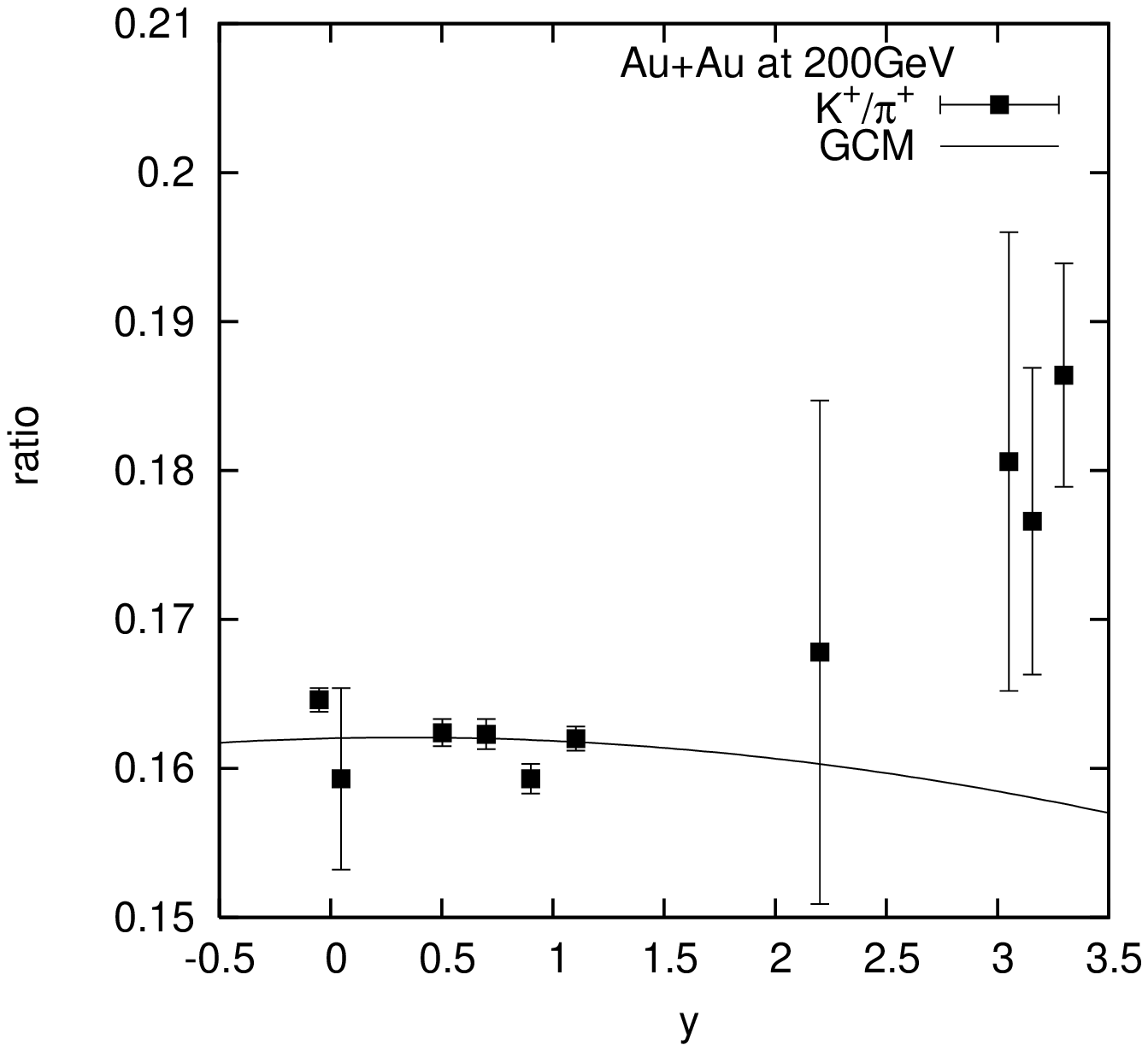}
\end{minipage}}%
\subfigure[]{
\begin{minipage}{.5\textwidth}
\centering
 \includegraphics[width=2.5in]{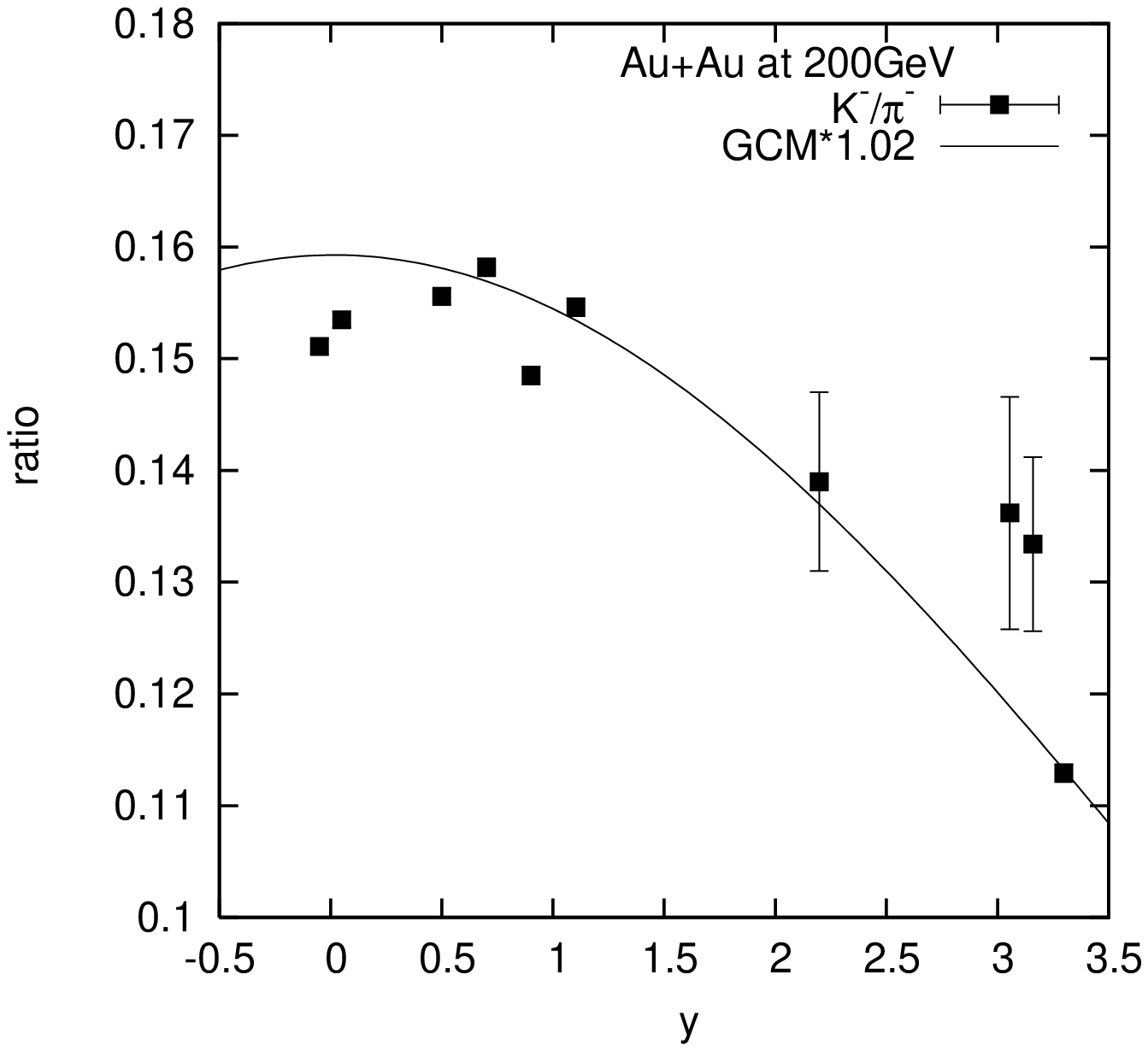}
 \end{minipage}}%
\caption{Full phase-space $K/\pi$ ratios as a function of rapidity systematics at $\sqrt{S_{NN}}$ = 200GeV. Errors are
statistical only. The different experimental points are from Ref.\cite{Bearden1} and the parameter values are taken
from Table5. The solid curves provide the GCM-based results.}
\end{figure}

\begin{figure}
\centering
\includegraphics[width=2.5in]{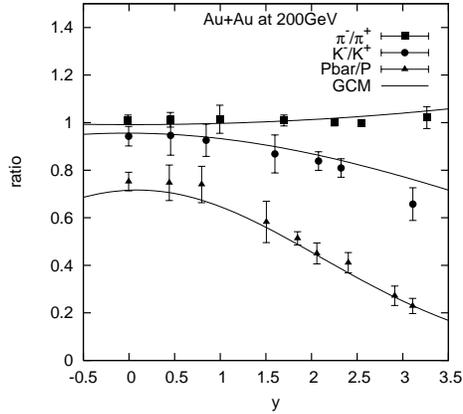}
\caption{Antiparticle-to-particle ratios as a function of rapidity
in central Au+Au collisions at $\sqrt{S_{NN}}$ = 200 GeV. The solid
lines are our results for $\pi^-/\pi^+$, $K^-/K^+$, $Pbar/P$
respectively. The experimental data are from Ref.\cite{Song1} and
the parameter values are taken from Table5.}
\end{figure}

\begin{figure}
\subfigure[]{
\begin{minipage}{.5\textwidth}
\centering
\includegraphics[width=2.5in]{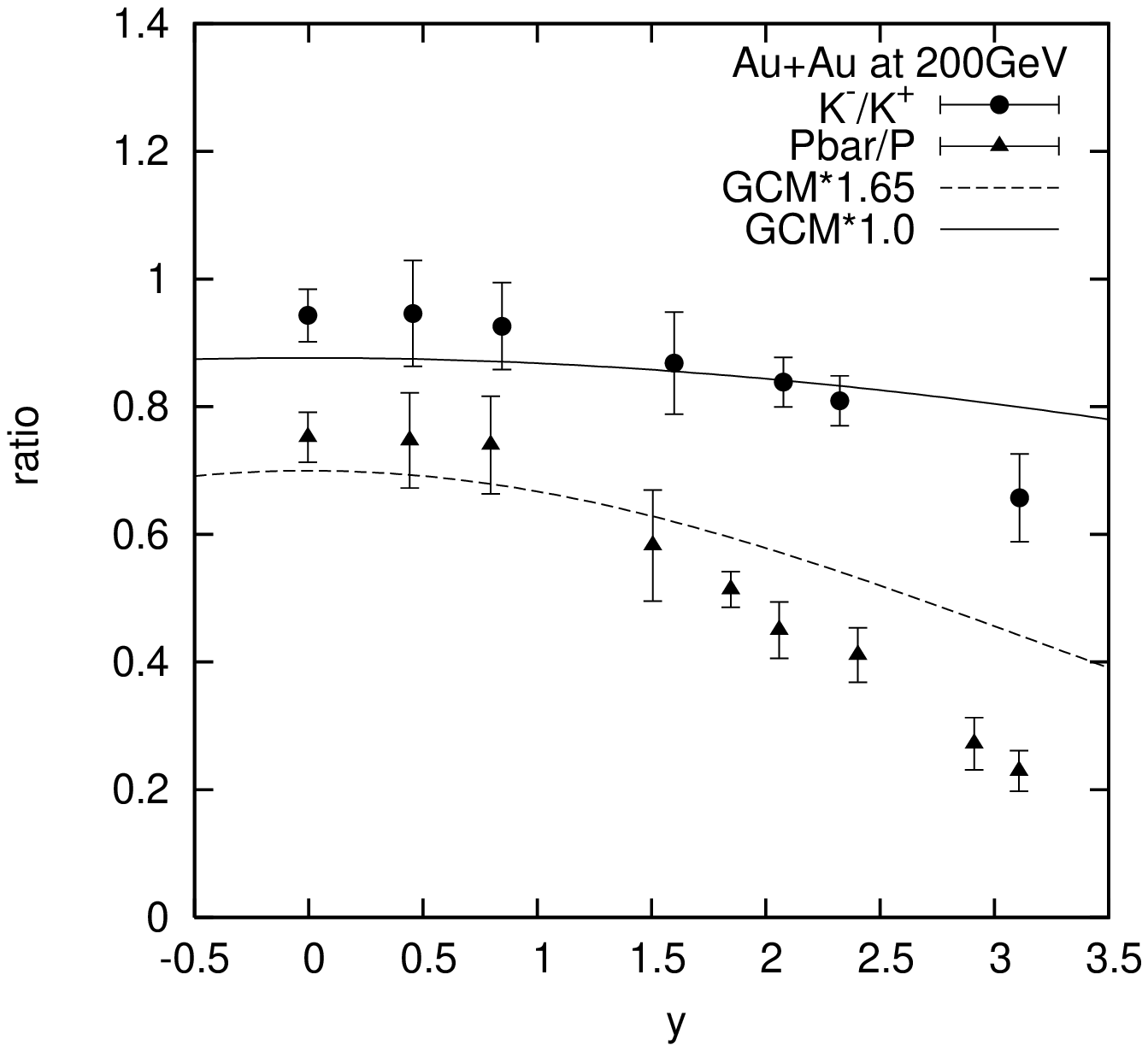}
\end{minipage}}%
\subfigure[]{
\begin{minipage}{.5\textwidth}
\centering
 \includegraphics[width=2.5in]{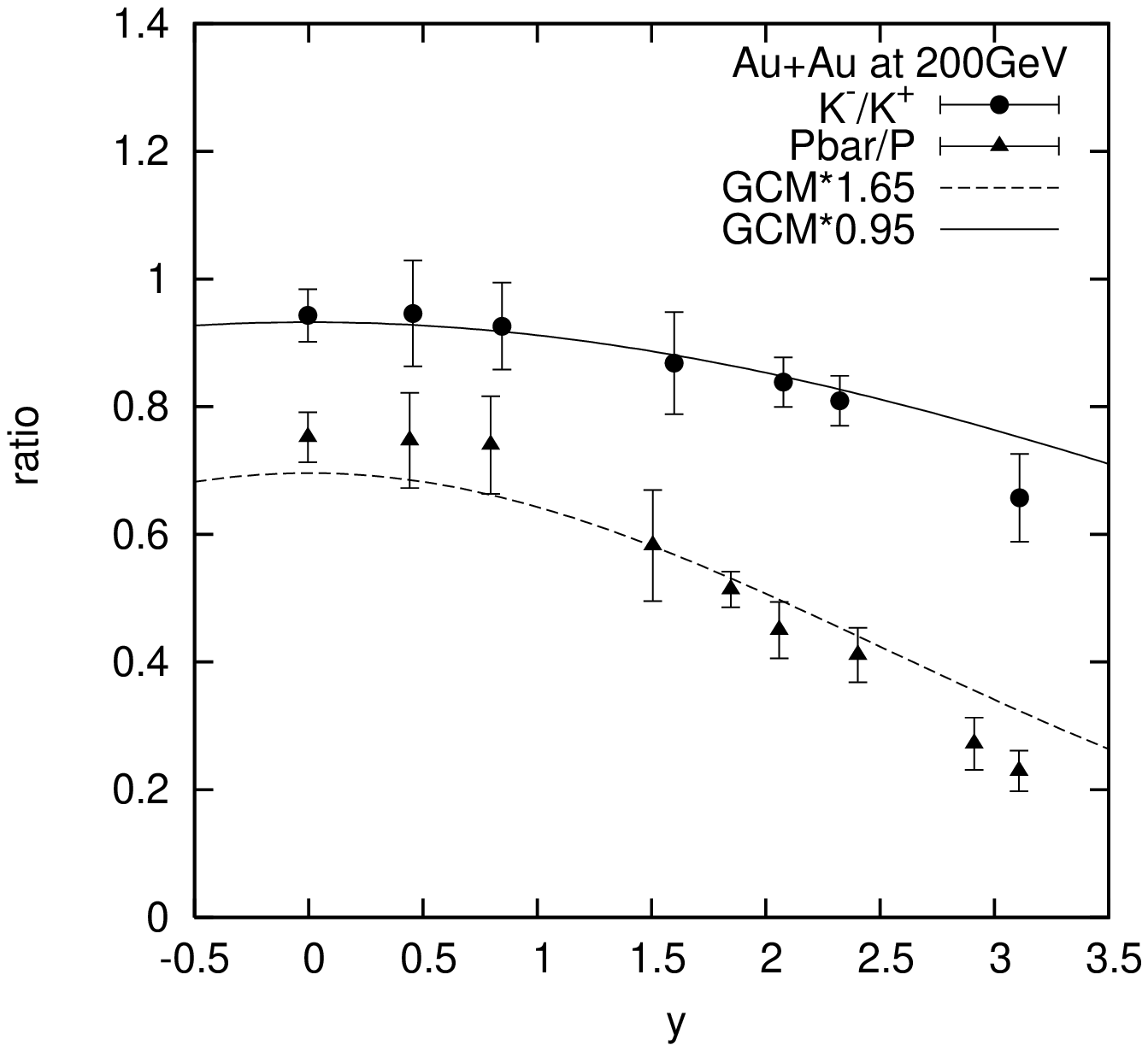}
 \end{minipage}}%
\caption{Antiparticle-to-particle ratios as a function of rapidity
in central Au+Au collisions at $\sqrt{S_{NN}}$ = 200 GeV. The solid
lines are our results for $K^-/K^+$ and $Pbar/P$. The experimental data are from Ref.\cite{Song1} and
the parameter values are taken from Table8 and Table9, taking $\Delta$=1.7 and $\Delta$=3.5
for Fig.10(a) and Fig.10(b) respectively.}
\end{figure}

\end{document}